\documentclass[acmsmall]{acmart}
\AtBeginDocument{%
  }

\setcopyright{acmcopyright}
\copyrightyear{2024}
\acmYear{2024}
\acmDOI{XXXXXXX.XXXXXXX}

\acmConference[CSCW'24]{Make sure to enter the correct
  conference title from your rights confirmation emai}{November 9--13, 2024}{San Jose, Costa Rica}
\acmISBN{978-1-4503-XXXX-X/18/06}

\usepackage{wrapfig}
\usepackage{paralist}
\usepackage{multirow}
\usepackage{bbding}
\usepackage{array}
\usepackage{graphicx}

\usepackage{censor}[2021-12-23]
\usepackage{keyval,stackengine,xparse,tikzpagenodes,tikz,soulpos}

\usetikzlibrary{calc,positioning}

\begin{document}

\title[BPCoach: Exploring Hero Drafting]{BPCoach: Exploring Hero Drafting in Professional MOBA Tournaments via Visual Analytics}

\author{Shiyi Liu}
\email{liushy@shanghaitech.edu.cn}
\affiliation{%
  \institution{School of Information Science and Technology, ShanghaiTech University}
  \city{Shanghai}
  \country{China}
}
\author{Ruofei Ma}
\email{marf@shanghaitech.edu.cn}
\affiliation{%
  \institution{School of Information Science and Technology, ShanghaiTech University}
  \city{Shanghai}
  \country{China}
}
\author{Chuyi Zhao}
\email{czhao07@tufts.edu}
\affiliation{%
  \institution{Department of Computer Science, Tufts University}
  \city{Medford}
  \state{Massachusetts}
  \country{USA}
}

\author{Zhenbang Li}
\email{lizhb@shanghaitech.edu.cn}
\affiliation{%
  \institution{School of Information Science and Technology, ShanghaiTech University}
  \city{Shanghai}
  \country{China}
}

\author{Jianpeng Xiao}
\email{xiaojianpeng@52tt.com}
\affiliation{%
  \institution{Guangzhou Qucheng Culture Media Co.}
  \city{Guangzhou}
  \country{China}
}

\author{Quan Li}
\email{liquan@shanghaitech.edu.cn}
\affiliation{%
  \institution{School of Information Science and Technology, ShanghaiTech University, and Shanghai Engineering Research Center of Intelligent Vision and Imaging}
  \city{Shanghai}
  \country{China}
}

\renewcommand{\shortauthors}{Liu, et al.}
\begin{abstract}
Hero drafting for multiplayer online arena (MOBA) games is crucial because drafting directly affects the outcome of a match. Both sides take turns to ``ban''/``pick'' a hero from a roster of approximately $100$ heroes to assemble their drafting. In professional tournaments, the process becomes more complex as teams are not allowed to pick heroes used in the previous rounds with the ``best-of-$N$'' rule. Additionally, human factors including the team's familiarity with drafting and play styles are overlooked by previous studies. Meanwhile, the huge impact of patch iteration on drafting strengths in the professional tournament is of concern. To this end, we propose a visual analytics system, \textit{BPCoach}, to facilitate hero drafting planning by comparing various drafting through recommendations and predictions and distilling relevant human and in-game factors. Two case studies, expert feedback, and a user study suggest that \textit{BPCoach} helps determine hero drafting in a rounded and efficient manner.
\end{abstract}



\maketitle
\raggedbottom
\section{Introduction}
\label{sec:intro}
Multiplayer Online Battle Arenas (MOBAs) have emerged as one of the most popular branches of esports. Prominent titles such as \textit{Defense of the Ancients 2 (DotA2)}, \textit{League of Legends (LoL)}, and \textit{Honor of King (HoK)} have successfully organized professional tournaments on an international scale, reaching millions of global viewers through live streaming~\cite{Li2018-wv,Kokkinakis2020-uo}.

\par In a standard match, two teams consisting of five players each engage in a battle to destroy the opponent's base and emerge victorious. Each player assumes control of a character within the game, commonly referred to as a \textit{hero}, and collaborates with their teammates to launch attacks against the opposing team. The drafting phase is a strategic stage at the beginning of the game where players select their respective heroes, taking into account their abilities and how they synergize with the team's overall strategy. Previous studies have demonstrated that the drafting phase significantly influences the outcome of a game~\cite{Semenov2017-gr,Kim2016-cy}. Furthermore, these games offer a vast selection of over 100 heroes, resulting in an astonishing number of approximately $1.56\times 10^{16}$ possible combinations~\cite{Hanke2017-xi}.

\par An advanced drafting tool has the potential to increase the strategic complexity of the game, leading to improved tactical decision-making by coaches and players and fostering better team cooperation. The sequence order of drafting carries great significance as it determines priority picks and facilitates effective counterpicking, allowing teams to leverage hidden information. Similar to the sequential drafting problems in other domains, such as the drafting process in the National Football League{~\cite{Berri2011-ud}} or the National Basketball Association{~\cite{Staw1995-oq}}, team formation{~\cite{Freeman2019-cs}}, and fantasy football game{~\cite{Lee2022-cp}}, it influences which team gains access to highly coveted heroes, enables strategic responses to the opponent's selections, and introduces an element of uncertainty for the opposing team. These enhancements can lead to a deeper understanding of the game's complexities, ultimately leading to increased player satisfaction and engagement.

\par Drafting in professional MOBA games plays a pivotal role, not only in terms of its strategic complexity, countering the enemy, and fostering team synergy, but also due to its direct impact on audience engagement and the broader game community. From the perspective of the audience, a more intricate drafting phase can generate thrilling and unpredictable matches, enhancing viewers' comprehension and appreciation of the game's strategic aspects, thereby enriching their overall viewing experience{~\cite{Tang2022-pv}}. Moreover, within the gaming community, esports transcends mere entertainment to become a global cultural phenomenon. Innovations like improved drafting tools can drive growth and maturity within the esports industry, leading to economic and employment opportunities{~\cite{Li2018-wv}}. These enhancements can also shape social dynamics and cultural values among gamers and fans, emphasizing the profound impact of these games within the community.

\par Despite the considerable research efforts dedicated to drafting recommendations or predictions in MOBA games~\cite{Hanke2017-xi,Summerville2016-gm,Chen2018-sl,Yu2019-jk,Chen2021-ls}, there are still several challenges when it comes to supporting professional matches, as the hero drafting in a professional tournament differs from regular games~\cite{Gourdeau2021-wy}. The first challenge arises from the ``Ban/Pick'' phase within the game itself. In standard tournaments, the best-of-$N$ rule is typically followed, where up to $N$ rounds are played, and the team that wins $(N+1)/2$ rounds is declared the winner~\cite{Chen2021-ls}. In each round, the ten players collectively choose heroes from a pool of over 100 candidates, each with distinct abilities, roles, and tactical styles~\cite{Chen2018-sl}. Furthermore, there are complex synergies and antagonistic relationships between these heroes~\cite{Kim2016-cy,Pobiedina2013-aj}. The ten heroes cannot be duplicated and are selected in a fixed sequence. Additionally, teams are not allowed to pick heroes that have been chosen in previous rounds due to the global ``Ban/Pick'' rule. Traditional recommendation systems are often built based on the single-round ban/pick rule~\cite{Chen2018-sl,Hanke2017-xi} or consider only the ``pick'' phase without the ``ban'' phase~\cite{Chen2021-ls}. The second challenge lies in the specificities of individual opponents. Existing studies treat all scenarios equally, disregarding the human factors present in matches~\cite{OpenAI2019-dc,Mohammed2022-ey,Gourdeau2021-wy,Porokhnenko2019-bx,Hanke2017-xi,Chen2018-sl,Chen2021-ls}. However, in professional matches, the particularities of a team or player, including their familiarity with drafting, tactical style, and individual abilities, heavily influence their decision-making and the match outcome. For instance, if an opponent excels with a particular hero, a team should prioritize preventing them from selecting that hero, even if it may not be necessary in most cases. Familiarity with heroes is crucial in MOBA games~\cite{Chen2018-kc,Yang2021-bf}. Players tend to pick heroes they are confident with while banning heroes their opponents excel at. Lastly, the frequent in-game updates and patch iterations continue to impact drafting strategies~\cite{Zhong2021-uc}. Minor updates may include slight changes to hero skills or health caps, while patch iterations can introduce significant alterations to game mechanics. Consequently, these changes have varying impacts on hero drafting~\cite{Wang2020-gh}. If a hero becomes weaker or less favorable due to updates, the necessity of selecting that hero diminishes. Since previous approaches utilize all available historical data during training, they cannot effectively support real-world applications due to dataset bias.

\par To address these research gaps, we have formulated and explored five research questions (\textbf{RQ1 -- RQ5}). Firstly, in order to investigate the actual requirements of coaches when making drafting decisions in professional games, we propose two questions: \textbf{RQ1: What is the information needed for a coach's decision on hero drafting?} and \textbf{RQ2: What should be an acceptable approach for information processing?} To answer \textbf{RQ1}, we conducted expert interviews with three collaborating experts. Through these interviews, we gained a valuable understanding of the coaches' needs. We considered hero drafting as a two-player zero-sum game, where the advantage gained by one player is equal to the disadvantage incurred by the other player. Based on the feedback received, we identified seven factors and derived five design requirements for the system (Section \ref{sec:designrequirement}) to address \textbf{RQ2}. We organized a design workshop to discuss potential interface designs for coaches. As a result, we developed a visualization system called \textit{BPCoach}, which provides recommendations and predictions for coaches in professional tournaments to aid in match preparation and hero drafting decisions. On the backend of the system, we employed Monte Carlo Tree Search (MCTS) to calculate the optimal next pick for our side. On the front end, we provided visualizations and interactions that present optimal and alternative drafting paths, specific team or player data, hero data, and in-game change data. Furthermore, we incorporated carefully designed interactions and auxiliary views to facilitate users in exploring information and making favorable decisions.

\par Based on the designed system prototype, we further investigated three research questions: \textbf{RQ3: What scenarios and when will this system be used?} \textbf{RQ4: What is the usability and effectiveness of the adaptive system in hero drafting?} and \textbf{RQ5: How will coaches interact and collaborate with such an AI-enabled system?} To address these questions, we explored two distinct usage scenarios, conducted expert interviews, and performed a user study. The two usage scenarios exemplify the practical application of the system. Experts utilized the system to iteratively interact and refine their decisions. During the expert interviews, we engaged in discussions with experts, covering topics such as system performance, visual design, interaction, generalizability, and scalability. Additionally, we conducted a within-subjects user study involving 12 participants, including three professional gamers and nine students with extensive gaming experience from a local university. The results of the study revealed that the proposed system demonstrated superiority in terms of informativeness, decision-making support, and system usability. In summary, the contributions of this work are as follows:

\begin{compactitem}
\item We have proposed a hero drafting approach that is based on the combination of MCTS, Markov chains, and Random Forest (RF). This approach is specifically designed to support the global ``ban''/``pick'' rules observed in best-of-$N$ tournaments.
\item We proposed an interactive visual analytics system that aids users in developing a comprehensive understanding of the game and making well-considered drafting decisions. This system visualizes the human factors of specific opponents and the impact of in-game changes, enabling users to make informed decisions during the drafting process.
\item Our system's efficacy has been validated through two usage scenarios, expert feedback, and a user study. Furthermore, we conducted an evaluation comparing our new model with three baseline models to showcase the superiority and effectiveness of our approach.
\end{compactitem}

\section{Hero Drafting in \textit{King Pro Leagues}}
This work was developed and evaluated based on professional tournaments such as \textit{Hok}, \textit{King Pro Leagues (KPL)}~\cite{TencentKPL}. These tournaments adhere to the best-of-$N$ rule and follow the global `ban''/``pick'' rule. The drafting process involves two teams taking turns to pick or ban heroes in the order of ``b1-b2-b1-b2-p1-p2-p2-p1-p1-p2-b2-b1-b2-b1-p2-p1-p1-p2'', where ``b''/``p'' represents ``ban''/``pick'' respectively, and ``$1$''/``$2$'' represents the team side. Furthermore, the ten heroes picked must be unique, and teams cannot select heroes that have already been picked in previous rounds~\cite{KplRule}. This necessitates a trade-off between the average strength over the $N$ rounds and the immediate strength in a single round.

\par The \textit{KPL} consists of approximately $20$ teams, with relatively stable player lineups. Each team possesses its own unique style and understanding of drafting strategies. Typically, teams attempt to anticipate their opponents' tendencies in order to limit their options and assert control over the game. Simultaneously, teams strive to strike a balance between hero relationships and their familiarity with those heroes. It's worth noting that \textit{HoK} experiences regular updates, ranging from minor tweaks to significant changes~\cite{KplUpdates}. The \textit{KPL} promptly incorporates these updates, as they can have a substantial impact on hero drafting. For instance, the hero ``\textit{Wang Zhaojun}'' was a popular pick with a win rate of over $50\%$ in spring 2022~\cite{Hero20200001}. However, after a patch iteration that shifted the game's focus towards combat, ``\textit{Wang Zhaojun}'' became less effective, resulting in a win rate drop to $34.9\%$ in the fall of 2022~\cite{Hero20200004}. Consequently, the number of times ``\textit{Wang Zhaojun}'' was picked also decreased by more than half.

\par An exemplary instance of a drafting victory in \textit{KPL} took place during a highly anticipated match between two renowned teams, \textit{GK} and \textit{eStarPro}. Playing on the red side, \textit{GK} skillfully set up a drafting trap. In the first ``ban'' stage, instead of opting for conventional bans like \textit{eStarPro} did, \textit{GK} made a bold move by banning the highly skilled heroes of their opponents, namely \textit{``Zhang Fei''} and \textit{``Zhuge Liang''. }  This put \textit{eStarPro} in a difficult situation as the three top-tier heroes at that time, namely \textit{``Jiang Ziya''}, \textit{``Pei Qinhu''}, and \textit{``Cheng Yaojin''}, were left open for selection. If \textit{eStarPro} chose one of these heroes, \textit{GK} would have the opportunity to pick the remaining two, resulting in a disastrous scenario for \textit{eStarPro}. After careful deliberation,  \textit{eStarPro} reluctantly picked \textit{``Jiang Ziya''}. Capitalizing on this, \textit{GK} seized the opportunity and selected \textit{``Pei Qinhu''} and \textit{``Cheng Yaojin''}, forming a formidable combination. In the second stage of the ban-pick phase, playing on the blue side, \textit{eStarPro} finalized their lineup first. \textit{GK} had the final pick and strategically chose the counter hero, \textit{``Nezha''}, who posed a significant threat. This proved to be troublesome for \textit{``Marco Polo''}, the hero chosen by \textit{eStarPro}. Additionally, their \textit{``Jiang Ziya''} in the mid lane is also an easy target for \textit{``Nezha''}. Moreover, \textit{``Nezha''} exhibited ideal early game damage, which complemented the overall transition of \textit{GK} in the early game. This final selection of \textit{``Nezha''} laid a solid foundation for the performance of \textit{GK} in the match. The strategic planning of \textit{GK} was meticulous and interconnected, with each move seamlessly leading to the next, showcasing their flawless calculations and execution.

\section{Related Work}
\subsection{Game Data Analysis and Visualization}
The rich data collected in  games have boomed the field of game data analysis, including game analytics~\cite{Kokkinakis2020-uo,Freeman2019-cs,Kow2013-hn,Grandprey-Shores2014-ra,Ducheneaut2007-qc} and machine learning~\cite{OpenAI2019-dc,Mohammed2022-ey,Gourdeau2021-wy,Porokhnenko2019-bx,Hanke2017-xi,Chen2018-sl,Chen2021-ls}. The main process of these efforts is to extract specific patterns in the game and then make some evaluations or predictions. Machine learning work includes win rate prediction~\cite{Kim2020-pg,Hodge2021-ni,Lee2020-ae}, highlight detection~\cite{Schubert2016-by}, factors and strategies affecting the outcome~\cite{Ahmad2019-uq,Xia2019-uj,Pobiedina2013-jf}, and recommendation systems~\cite{Najafabadi2019-pu,Bertens2018-go}.

\par Visualization of game data benefits multiple target groups, such as game developers~\cite{Li2017-dn, Li2021-nh, Wallner2021-fy} and non-professional players~\cite{Charleer2018-kv,Wallner2016-xe}. Game developers use visual analytics to detect imbalances in-game mechanics~\cite{Li2017-dn} and improve the player experience~\cite{Wallner2021-fy}. For example, Ducheneaut and Moore~\cite{Ducheneaut2004-iu} analyzed player-to-player interactions and concluded with some recommendations for the design and support of social activities within multiplayer games. Wallner et al.~\cite{Wallner2021-fy} focused on post-game data visualization to explore information needs raised by players. Javvaji et al.~\cite{Javvaji2020-eu} presented an interactive visualization based on game telemetry data for game designers to study player patterns. Player-centered visualizations can support and facilitate gameplay~\cite{Medler2011-kf}. Wallner et al.~\cite{Wallner2021-fy} built a training visualization that allows players to review their gameplay and thus learn from experience and from each other. Charleer et al.~\cite{Charleer2018-kv} designed a dashboard view that effectively facilitates user insight and experience. Afonso et al.~\cite{Afonso2019-of} employed visualization techniques on spatiotemporal data to allow players to better understand their strategies.

\par Previous work on data analysis and visualization of games has focused on in-game~\cite{Li2017-dn,Wallner2021-fy,Afonso2019-of} and post-game data~\cite{Wallner2021-fy}, facing both non-professional players and game designers. Numerous popular commercial tools{~\cite{Cook2022-rl}} such as \textit{OP.GG}{~\cite{OPGG}}, \textit{senpai}{~\cite{SenpAI}}, and others offer gamers access to historical data for reference during the drafting and provide various indicators related to heroes and players. These tools summarize information such as hero win rates, hero relationships, and individual player win rates. For instance, \textit{senpai} provides in-game assistance by generating hero selection suggestions for the next step and allows players to access information such as hero win rates, counter relationships, and player game status. However, most of these existing tools primarily cater to non-professional players, and as a result, they lack statistics on team-specific hero combinations and do not support the global ``ban''/``pick'' rules observed in professional games. Additionally, none of these tools evaluate how in-game changes affect heroes. Players often seek information on whether a hero's win rate has increased or decreased, or whether changes have made a hero more powerful. This study focuses on pre-game data analysis and visualization for professional teams. We designed an interactive visualization system that displays pre-game opponent team statistics in terms of heroes and players, to help teams visualize the situation and make thoughtful drafting decisions.

\subsection{Hero Drafting Recommendation Techniques}
Many academic works revolve around hero drafting recommendations, which fall into three main types: algorithms based on (1) prediction, (2) win rate, and (3) zero-sum game theory~\cite{Rapoport2013-yt}.

\par The prediction-based works~\cite{Summerville2016-ds,Summerville2016-gm,Yu2019-jk} treat the hero drafting problem as a sequential prediction problem. The output is the most likely choice, rather than the best choice. Win rate-based work focuses on picking heroes that bring higher win rates to the drafting. For example, Hanke et al.~\cite{Hanke2017-xi} developed a mechanism based on association rules that suggests more suitable heroes. Other works have explored different methods for win prediction, including natural language processing (NLP)-based methods~\cite{Mohammed2022-ey}, regression-based methods~\cite{Porokhnenko2019-bx} and neural network (NN)-based methods~\cite{Porokhnenko2019-bx, Gourdeau2021-wy} to recommend the best drafting. These works output locally optimal recommendations rather than globally optimal ones. In the third category, the researchers treat the drafting problem as a two-player game and apply the game theory. For example, \textit{OpenAI Five} uses the Minimax algorithm~\cite{Fan1953-zg} to select two 5-player teams from a total of 17 heroes. When about 100 heroes are involved, the computational complexity increases dramatically, which makes the method infeasible. Chen et al.~\cite{Chen2021-ls} first applied Monte Carlo Tree Search (MCTS) to the global ``ban''/``pick'' rule. In this work, the global evaluation of the reward in MCTS is less appropriate because it is the sum of $N$ round win rates. In this case, a very high win rate in one round and a low win rate in the rest of the rounds ($<50\%$) can be used as output, while the win condition is to win $(N+1)/2$ rounds. 

\par Nevertheless, none of the previous works considered the different tendencies of different human teams. Moreover, most of these works ignored ``ban'' games and focused more on ``pick''~\cite{Hanke2017-xi,Porokhnenko2019-bx,Chen2018-sl,Chen2021-ls}. To address these issues, we propose a model built on MCTS, Markov chains, and RF. We design a new evaluation principle in MCTS to satisfy the $N$ optimal winning condition. We also predict the possible choices using the Markov chain to reduce the computational complexity.

\subsection{Visualization of Sequence and Alternatives}
The visualization of drafting paths can be considered the problem of visualizing prediction sequences and the alternatives. Traditional sequence visualization can be divided into three categories, including timeline-based methods~\cite{Karam1994-sl}, tree-based methods~\cite{Monroe2013-wb}, and Sankey-based methods~\cite{Riehmann2005-cd}. Timeline-based approaches are the most intuitive methods, which organize events directly in chronological order, where some work aggregate events in various time granularities~\cite{Xie2020-cl}. In tree-based approaches, the sequences are arranged in a tree structure to show the hierarchy~\cite{Wongsuphasawat2011-ss,Arazy2015-gk}. Each node in the tree structure represents an event and is connected to adjacent events~\cite{Wongsuphasawat2014-cs,Monroe2013-wb,Du2017-uq}. Sankey-based approaches can better represent the transition relationships of events~\cite{Guo2021-gg}. Events are encoded as nodes of different widths according to their proportion in the timestamp~\cite{Wongsuphasawat2012-zt,Perer2013-fh,Riehmann2005-cd}.

\par Since a hero drafting path is a sequence of events in a fixed order, a straightforward approach is to place the events on a timeline, as in \textit{Lifelines}~\cite{Plaisant1998-dm} and \textit{CloudLines}~\cite{Krstajic2011-ix}. However, the timeline-based approach has difficulties in representing alternative paths. The traditional tree-like structure~\cite{Wongsuphasawat2014-cs,Qi2020-fp} does not take into account the weights of nodes with different scores or probabilities in the hero drafting sequence. Du et al.~\cite{Du2017-uq} used the interleaved heights of nodes to represent the weights. However, the game boasts a roster of around 100 heroes, which consequently creates roughly 100 branches to consider. Unfortunately, previous efforts have struggled to effectively address the resulting scalability issues. Guo et al.~\cite{Guo2019-fp} used a circular glyph with colored ring layers to visualize uncertainty and alternatives. The user gets the path by exploring from the inner ring, which makes the full picture of the path unintuitive. Therefore, we propose a new interactive visualization method that combines a modified tree structure with glyphs to distinguish top predictions from alternatives and to satisfy scalability and intuitiveness requirements.

\section{Formative Study}
\subsection{Need-Finding Interview}
The objective of this work is to provide professional teams with comprehensive advice and insights regarding pre-game hero drafting. To ensure the effectiveness and relevance of our work, we collaborated with three experts in hero drafting who helped us extract detailed requirements. E1, a training manager (male, age: $28$), has extensive experience and has been closely following the esports scene for several seasons. E2, an experienced coach (male, age: $30$), possesses two years of coaching experience and assumes the responsibility of directing hero drafting and game strategy. E2 is the key decision-maker when it comes to hero drafting for official matches. E3, a professional player (male, age: $19$), brings valuable insights from his vast experience of playing in hundreds of matches. In the team hierarchy, E1 is responsible for managing all aspects of the team, E2 leads hero drafting and game strategy, and E3 takes charge of the game while also providing his own ideas during hero drafting. 

\par When discussing the significance of hero drafting in professional tournaments, E2, the coach responsible for drafting decisions, emphasized that the ``ban''/``pick'' stage directly determines the strategic path that the team must rely on to secure victory. Once the drafting is completed, it is primarily a matter of executing the chosen strategy. According to E2, a correct strategic path is the foundation for successful tactical execution. From their perspective, drafting serves as the cornerstone for all subsequent strategies. Furthermore, E3, a player, highlighted that in high-end professional leagues where players possess similar levels of skill, gaining an advantage in hero drafting can make it easier to secure a victory. According to E3, when the coach provides an advantageous drafting scenario, it increases the team's chances of winning. E2 also mentioned a notable incident where a well-known coach utilized the global ``ban''/``pick'' rule to defeat a stronger team. This achievement instantly propelled the coach to celebrity status among game fans. This highlights that high-quality drafting is highly desired and appreciated by game enthusiasts, as it can greatly influence the outcome of a match and create an exciting experience for fans.

\par To ascertain the requirements for high-quality drafting, we conducted a 40-minute online interview with the experts mentioned earlier. During the interview, we initially inquired about the conventional approach and the factors that should be taken into account when making ``ban'' or ``pick'' decisions (\textbf{RQ1}). Generally, the current prevalent method of hero drafting relies on prior experience and gathering data on opponents beforehand. Experts used to consider the following factors: (\textbf{F1}) drafting sequence, (\textbf{F2}) player's performance with a specific hero, (\textbf{F3}) player's overall status, (\textbf{F4}) team performance, (\textbf{F5}) play style, (\textbf{F6}) hero's adaptability to in-game changes, and (\textbf{F7}) relationships among heroes. Meanwhile, experts suggest that only the most recent year or so of relevant data is meaningful, as there will be fewer significant changes in the game during this period.

\par Next, we inquired about the challenges associated with the traditional approach. When confronted with unexpected changes, E2 mentioned that ``\textit{considering the global 'ban'/'pick' rule, there are instances where I am unable to swiftly provide the optimal global solution, and every decision in the game is time-sensitive.}'' In the traditional method of gathering opponent information, they manually input the required data into \textit{Excel} and utilize its functions, such as averaging and sorting. E1 expressed, ``\textit{this intricate process lacks intuitiveness and interactivity.}'' Moreover, the organized data, primarily utilized for pre-match planning, prove to be of limited assistance to coaches during the game due to the time-consuming nature of acquiring it. Additionally, in-game adjustments often become apparent only after a hero has failed multiple times.

\subsection{Survey on Existing Pre-game Statistics Platforms}
\label{dis:comp}
\begin{table}[ht]
\caption{Comparisons with top commercial platforms, the official \textit{KPL} data website, and \textit{BPCoach} in terms of seven factors.}
\label{tab:comp}
\scalebox{0.58}{\begin{tabular}{@{}cccccccccc@{}}
\toprule
&\textit{mobalytics}~\cite{Mobalytics}&\textit{senpai}~\cite{SenpAI}&\textit{porofessor}~\cite{Porofessor}&\textit{blitz}~\cite{Blitz}&\textit{op.gg}~\cite{OPGG}&\textit{facecheck}~\cite{Facecheck}&\textit{u.gg}~\cite{UGG}&\textit{KPL}~\cite{TencentKPL}&\textit{\textbf{BPCoach}} \\ \midrule
\textbf{F1} drafting sequence                                 & \Checkmark & \Checkmark & ·          & ·          & ·          & \Checkmark & ·          & ·          & \Checkmark \\
\textbf{F2} player's performance with a specific hero           & ·          & \Checkmark & \Checkmark & ·          & \Checkmark & \Checkmark & \Checkmark & \Checkmark & \Checkmark \\
\textbf{F3} player's overall status                           & ·          & \Checkmark & \Checkmark & ·          & \Checkmark & \Checkmark & \Checkmark & \Checkmark & \Checkmark \\
\textbf{F4} team performance                                  & ·          & ·          & ·          & ·          & ·          & ·          & ·          & ·          & \Checkmark \\
\textbf{F5} play style                                        & ·          & \Checkmark & \Checkmark & ·          & ·          & \Checkmark & \Checkmark & ·          & \Checkmark \\
\textbf{F6} hero's adaptability to in-game changes                 & ·          & ·          & ·          & ·          & ·          & ·          & ·          & ·          & \Checkmark \\
\textbf{F7} relationships among heroes                         & \Checkmark & \Checkmark & ·          & \Checkmark & \Checkmark & \Checkmark & \Checkmark & ·          & \Checkmark \\\bottomrule
\end{tabular}}
\end{table}

With the advent of the digital age, gamers have become increasingly focused on data-driven approaches{~\cite{Kokkinakis2020-uo}}. Consequently, numerous platforms that provide performance statistics have emerged in the market. Since \textit{HoK} has limited commercial platforms available, we selected platforms that support \textit{LOL}, which shares a similar drafting mechanism with \textit{HoK}. In Table{~\ref{tab:comp}}, we conducted a comprehensive comparison between these leading commercial platforms, and the official \textit{KPL} data website, based on the seven factors suggested by our experts. Platforms like \textit{mobalytic}, \textit{senpai}, and \textit{facecheck}, are capable of generating drafting recommendations based on the current drafting sequence (\textbf{F1}). However, these commercial platforms cater to non-professional players and lack compatibility with the global ``ban''/``pick'' rule. Only the official \textit{KPL} data website records this specific statistic. Some platforms, such as \textit{Senpai} and \textit{Porofessor}, offer hero recommendations during drafting, but they solely focus on suggesting heroes based on synergistic relationships without considering the overall strength of the complete lineup that both teams will eventually have. Most platforms provide the player's performance with a specific hero (\textbf{F2}), the player's overall status (\textbf{F3}), and the relationships among heroes (\textbf{F7}). However, no platform consolidates the usage of specific hero combinations by particular teams (\textbf{F4}), even though non-professional players often prefer fixed combinations of heroes who rank together. This is likely due to the complexity involved in quantifying team collaboration indicators, which encompass multiple dimensions such as the players themselves, the chosen heroes, and the combinations of heroes selected by the players. Regarding play styles (\textbf{F5}), many platforms such as \textit{senpai} and \textit{Porofessor}, \textit{facecheck}, and \textit{u.gg}, assign labels to players. These labels are typically determined by algorithms based on in-game data, selecting a suitable label from a limited pool. While these labels provide some information, they lack personalization. Lastly, all platforms offer real-time statistics that are regularly updated. However, users often find it challenging to remember every statistic at any given time, making it difficult to identify trends through clear comparisons (\textbf{F6}). In the subsequent section (Section{~\ref{dis:realize}}), we will discuss how our proposed system \textit{BPCoach} extracts these factors and contributes to the improvement of pre-game data platforms.
\subsection{Design Requirements}
\label{sec:designrequirement}
In order to ensure that our approach aligns well with the tasks and requirements of the domain, we further consulted all experts (E1-E3) regarding their main concerns and design requirements for the demo system (\textbf{RQ2}). Based on their feedback, we have summarized a list of design requirements as follows:

\begin{compactitem}
\item \textbf{DR1. Recommend and predict the ``ban''/``pick'' option.} Both E1 and E2 expressed their need for recommended ``ban''/``pan'' choices for their own team. While the general hero drafting is typically determined by the coach's experience, they wanted a limited set of heroes recommended based on objective winning data from a pool of over $100$ choices. Additionally, these recommendations should align with the global ``ban''/``pick'' rules in a best-of-$N$ tournament. Furthermore, all three experts agreed that the ability to predict the opponent's choices would be valuable in a hero drafting game. For instance, if teams notice a trend of opponents favoring powerful heroes, they can strategically ``ban'' or ``pick'' those heroes or make early choices to prevent their opponents from gaining a strong drafting advantage. Therefore, predictive capabilities can assist in countering opponent strategies.

\item \textbf{DR2. Display hero drafting paths and alternatives.} E2 and E3 emphasized the importance of visualizing the hero drafting path, which includes iterative recommendations and predictions. Such visual representation would provide a comprehensive overview of the entire drafting process, facilitating better decision-making. Additionally, alternative paths should be presented to offer a range of references and enable specific decisions based on other relevant factors.

\item \textbf{DR3. Support ``what-if'' assumptions and predict win rates.}  In addition to recommended or predicted choices, E2 and E3 expressed a desire for a ``what-if'' assumption feature. This functionality would enable them to make assumptions about different hero drafting scenarios and observe the corresponding win rates, thereby enhancing their understanding of the drafting process.

\item \textbf{DR4. Quantify the synergistic and antagonistic relationships between heroes.} E2 and E3 noted that while professional coaches and players generally have a sense of the synergistic and antagonistic relationships between heroes, these concepts remain abstract and unquantified. E2 mentioned, ``\textit{When we compare heroes, we can have a rough idea of their rating.}'' However, experts' subjective perceptions of these relationships vary. In cases where multiple heroes may be favorable choices but only one can be selected, quantifying and displaying these relationships would be valuable.

\item \textbf{DR5. Display data for specific teams and players.} All three experts highlighted the significance of making decisions based on specific opponents. For example, if an opponent's mid-laner consistently performs exceptionally well, the drafting strategy may involve targeting that player and banning their preferred heroes. Currently, teams manually extract data for specific opponents using tools like \textit{Excel}, which is cumbersome and unintuitive. Displaying team-specific and player-specific data would assist teams in effectively evaluating their opponents' status.

\item \textbf{DR6. Offer in-game updates and effects on heroes.} Both E2 and E3 mentioned the frequent updates in the game that have an impact on hero drafting. Understanding and adapting to these changes quickly can be challenging. By providing visual comparisons, they felt that they would be alerted to these updates and be able to swiftly comprehend the situation, enabling them to make informed decisions and trade-offs.
\end{compactitem}

\section{BPCoach}
\subsection{Data Description}
Based on the identified factors (\textbf{F1-F7}), we extract the following data in the system: 1) \textit{drafting sequence (\textbf{F1}).} The ``ban''/``pick'' heroes and their orders for both sides. 2) \textit{Player performance (\textbf{F2}).} Five player performance indicators, including \textit{KDA ratio} ($(Kill+Assist)/Death$), \textit{average damage per minute}, \textit{average damage taken per minute}, \textit{average cash gained per minute}, and \textit{participation}; 3) \textit{Player status (\textbf{F3})}. The number of victories or defeats overall or in some specific heroes.  4) \textit{Team performance (\textbf{F4}).} The performance of the team, including: \textit{win rate}, \textit{team KDA ratio}, \textit{average number of tyrants}, \textit{average number of dragons}, \textit{average number of towers destroyed}, and \textit{average game duration}; 5) \textit{Keywords about the team or player (\textbf{F5}).} Keywords extracted from news around the player or team, reflecting common heroes and playstyles; 6) \textit{Hero data (\textbf{F6}).} Data reflecting how heroes are drafted, such as win rate, picked rate, banned rate, etc. 7) \textit{Relationships between heroes (\textbf{F7}).} The winning yields the most synergistic heroes and the most antagonistic heroes.

\par The dataset used in our study is publicly available on the official \textit{KPL} data website~\cite{TencentKPL}. It comprises a total of $1,182$ professional battles that took place in the year 2022. This dataset includes information on hero drafting, drafting sequences, as well as player and team performance indicators. In addition to the battle data, we also collected a set of $6258$ \textit{KPL}-related news articles from the official news site~\cite{KPLNews}. These news articles primarily cover the content and outcomes of the games, providing insights into various aspects of teams and players. For instance, if a player delivers an outstanding performance with a specific hero, there will likely be news articles highlighting their achievements. By analyzing the frequency of appearances of certain heroes in the news articles associated with a particular player, we can infer that the player has been performing well with that hero.

\subsection{System Overview}
\begin{figure}[ht]
  \centering
  \includegraphics[width=\linewidth]{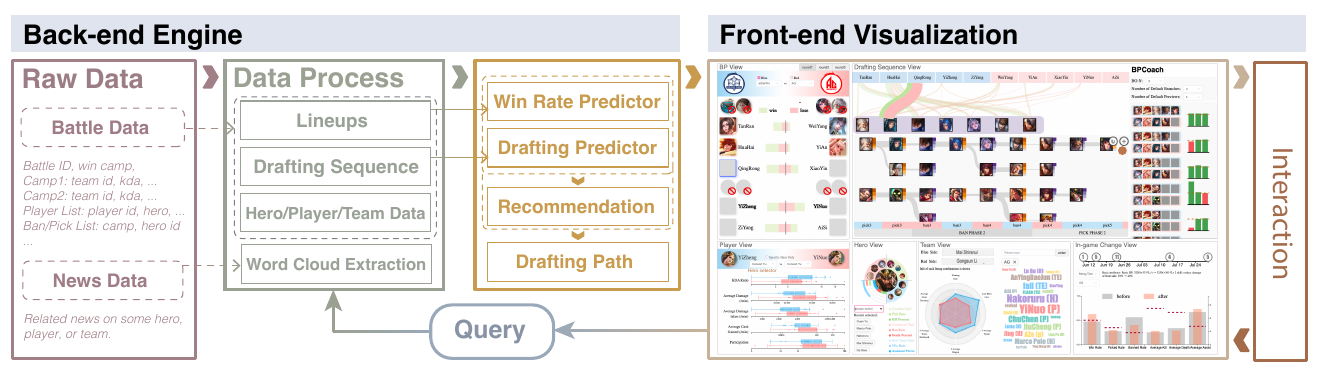}
  \caption{\textit{BPCoach} Pipeline: (1) Back-end engine to generate the recommendations and predictions; (2) Front-end visualization facilitates exploratory data analysis.}
  \label{fig:pipeline}
  \Description{}
\end{figure}

Based on the given requirements, we have developed a visual analytics system called \textit{BPCoach} (\textbf{RQ2}). Its purpose is to present the recommended hero drafting path and associated human factors, supporting professional teams during the drafting phase of games. The system architecture and pipeline are illustrated in Fig.~\ref{fig:pipeline}. To begin, we collect and process battle logs to extract drafting information, including hero drafting sequences, as well as specific data related to heroes, players, and teams for both sides. The extracted drafting information is then utilized to train the win rate predictor (\textbf{DR4}). Simultaneously, the drafting sequences are used to construct the drafting predictor model (\textbf{DR2}). The recommendations for hero selection are generated based on the win rate predictor and the drafting predictor (\textbf{DR1}). The iterative process of recommendations and predictions forms the drafting paths (\textbf{DR3}). To facilitate the exploration and comprehension of the data, \textit{BPCoach} incorporates interactive visualizations. The user interface encompasses five primary views: \textit{bp view} (\textbf{DR5}), \textit{drafting sequence view} (\textbf{DR1--3}), \textit{player view} (\textbf{DR5}), \textit{hero view} (\textbf{DR4}), \textit{team view} (\textbf{DR5}), \textit{in-game change view} (\textbf{DR6}).

\subsection{Back-end Engine}
\begin{figure}[ht]
  \centering
  \includegraphics[width=\linewidth]{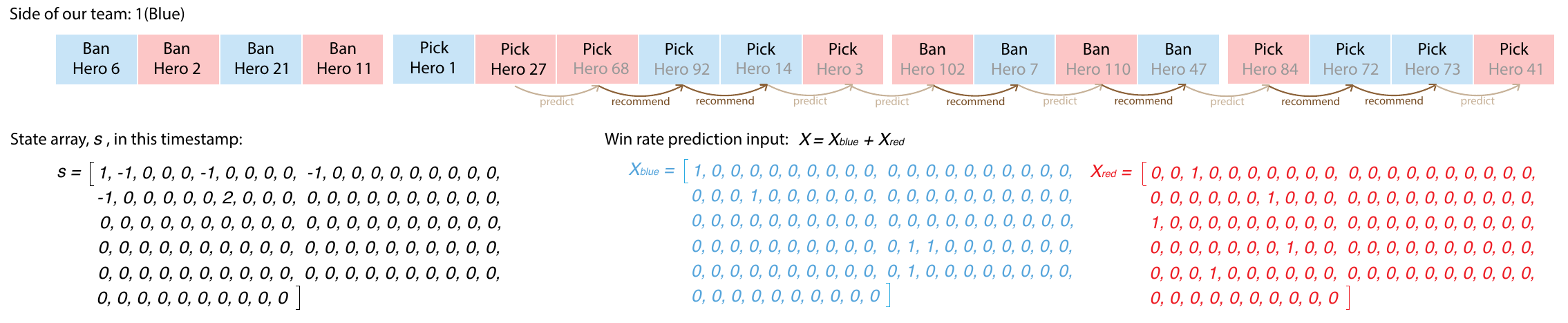}
  \caption{Recommendation or prediction iterations in hero drafting path and the corresponding state array and win rate predictor input.}
  \label{fig:banpick}
  \Description{}
\end{figure}

The back-end engine plays a crucial role in the BPCoach system. Its primary responsibilities include processing the raw data, executing the model, and transmitting the output to the front-end interface.

\subsubsection{Drafting Prediction}
We achieve drafting prediction utilizing the Markov chain, which offers both statistical significance and ease of comprehension~\cite{Norris1998-zf}. We define the terms as follows:
\begin{compactitem}
    \item \textbf{Team: } the two teams, $T$, in a best-of-$N$ match. We define $1$ as our side and $2$ as the opponent's side.
    \item \textbf{Team side:} blue or red in each round. We define $1$ as the blue side and $2$ as the red side.
    \item \textbf{State:} An array, $s$, with $110$ dimensions, indicating the current drafting state. As shown in Fig.~\ref{fig:banpick}, if a hero is banned, $s_{hero\_ID} = -1$. If a hero is picked by team $T$, $s_{hero\_ID} = T$.
    \item \textbf{Previous selection:} two arrays with $110$ dimensions, $Pr_{team}$ indicating heroes that have been used by this team or the opposing team in previous rounds. If $Pr_{1}[hero\_ID] = 1$, the hero cannot be selected by team 1.
\end{compactitem}

\par To generate the probability of a state change from $s$ to $s'$, i.e., $P_{ss'}$, we calculate the number of transitions $N_{ss'}$ from $s$ to $s'$ within the given drafting sequence. By calculating the ratio of $N_{ss'}$ to $N_{s}$, we can obtain $P_{ss'}$. Thus, the transition probability matrix $P$ is generated to predict the drafting sequences. After possible heroes for team $T$ are generated, if the hero is picked in previous rounds, $Pr_{T}[hero\_ID] = 1$, the possibility to select this hero in the matrix $P$ is assigned as $0$. Based on this $P$, we provide the top three predictions to the user, and the most probable ``ban''/``pick'' choice is identified as \textbf{top prediction}.

\begin{wraptable}{r}{4.5cm}
\caption{Performance of team win rate predictor.}
\label{table:performance}
\setlength{\tabcolsep}{3mm}{
\begin{tabular}{@{}ccc@{}}
\toprule
     & Accuracy   & AUC   \\ \midrule
LR   & 0.528 & 0.532 \\
RF   & \textbf{0.564} & \textbf{0.572} \\
GBDT & 0.523 & 0.523 \\
NN   & 0.507 & 0.513 \\ \bottomrule
\end{tabular}}
\end{wraptable}

\subsubsection{Win Rate Predictor and Comparison with Baseline Models}

We tested four models (RF~\cite{Pal2005-ls}, NN~\cite{Bishop2006-zl}, Logistic Regression (LR)~\cite{Wright1995-dm}, Gradient Boosted Decision Tree (GBDT)~\cite{Ye2009-jn}) in win rate prediction and compared the accuracy and area under ROC curve (AUC). The dataset was randomly divided into training and testing sets ($8:2$). Blue and red drafting are encoded as two feature vectors of length $110$ and the input is the concatenation of the two vectors as shown in Fig.~\ref{fig:banpick}. The results are listed in Table~\ref{table:performance}. LR models the probabilities as linear combinations of individual features without awareness of the interactions between features and in turn performs relatively poorly. Both RF and GBDT are decision tree-based models. RF performs better as GBDT is more sensitive to outliers~\cite{Yang2016-jd}, and there is a good chance that a player's performance in a tournament will lead to unexpected results. Considering the small number of professional battles and the fact that NN is a complex model, it leads to over-fitting problems. Due to game balance and in-game uncertainty, in previous work, most winning predictors were accurate around $0.6$ at the cost of $100,000$ orders of magnitude data~\cite{Chen2021-ls,Chen2018-sl}. Therefore, the performance of RF trained with a limited number of professional games is acceptable. We thus choose the RF model as our winning predictor.

\subsubsection{Drafting Recommendation Model and Comparison with Baseline Models}
To satisfy the global ``ban''/``pick'' rule, we consider this process as a zero-sum two-player game. MCTS has been applied to many classic zero-sum games such as \textit{Go}~\cite{Silver2017-yb} and other real-time strategy games~\cite{Balla2009-wz}. Particularly, MCTS~\cite{Kocsis2006-sp,Coulom2007-ly} is a class of heuristic search algorithms to make decisions without building a full search tree. It expands the branches towards the most promising actions. The search tree is continuously expanded until the allocated time or memory resources are exhausted.
\begin{figure}[ht]
  \centering
  \includegraphics[width=\linewidth]{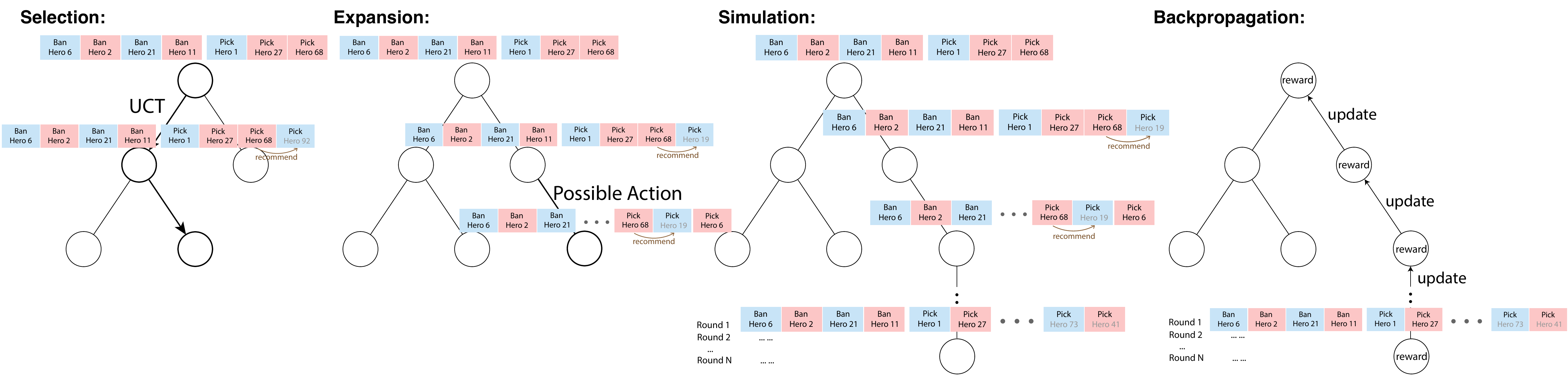}
  \caption{Four stages in MCTS.}
  \label{fig:mcts}
  \Description{}
\end{figure}
\par Drafting with a global ``ban''/``pick'' rule can be identified as a search tree with the maximum depth of $10 \times N$. Recommending a hero to our side is to find out the hero that leads to a global win while deciding which hero to ``ban'' is actually to find out which hero the opponent is most likely to win. Based on this logic, the MCTS in the draft game is built with the following four stages, as shown in Fig.~\ref{fig:mcts}:

\par \textbf{1) Selection.} The algorithm starts at the root node $R$ and successively selects child nodes until it reaches the leaf node $L$. The selection follows the Upper Confidence Bound (UCT) criterion~\cite{Auer2002-sx}: 
\begin{equation}
    A_t = \arg \max_{a} \bigg(Q_{t} (a) + c \sqrt{\frac{\ln (t)}{N_t(a)}}\bigg),
\end{equation}
where $Q_{t} (a)$ represents the current estimate of action $a$ at $t$, and $t$ represents the timesteps, $N_t(a)$ represents the number of times action $a$ is taken, and $c$ is the exploration term. The tree tends to explore nodes with higher estimates.

\par \textbf{2) Expanding.} Before reaching a leaf node, one of the potential actions is applied randomly to that node. To reduce the search time, we list all possible actions using the previous Markov chain prediction model. These actions are considered potential actions for the node to expand.

\par \textbf{3) Simulation.} Upon expanding a new node, the algorithm swiftly conducts random samplings to simulate the remainder of the game, adhering to the fundamental rules throughout the process.

\par \textbf{4) Back-propagation.} The reward of each node is backpropagated from the child nodes to the root node. 

\begin{wraptable}{r}{6cm}
\caption{Performance of MCTS hero drafting.}
\label{table:mcts}
\begin{tabular}{@{}ccc@{}}
\toprule
     & win rate in Bo3 & win rate in Bo5 \\ \midrule
RD   & 0.499$\pm$0.058       & 0.478$\pm$0.043       \\
HWR  & 0.732$\pm$0.053       & 0.658$\pm$0.034       \\
MCTS & \textbf{0.875$\pm$0.068}       & \textbf{0.830$\pm$0.075}       \\ \bottomrule
\end{tabular}
\end{wraptable}
We define the reward as the number of wins in the remaining rounds.
We use a win rate predictor to predict the outcome. The node's information is updated after each exploration.

\par After the number of iterations in these four steps above, the child node with the highest reward is returned as the output. Performance is evaluated as the win rate of model-selected drafting versus randomly-selected drafting in the ``best-of-3'' and ``best-of-5'' matches. We compare our MCTS with the random (RD) method and the highest win rate (HWR) method, where heroes are greedily selected based on the win rate of the round. Results with $95\%$ confidence intervals are listed in Tabel~\ref{table:mcts}.

\subsection{Front-end Visualization}
\begin{figure}[ht]
  \centering
  \includegraphics[width=\linewidth]{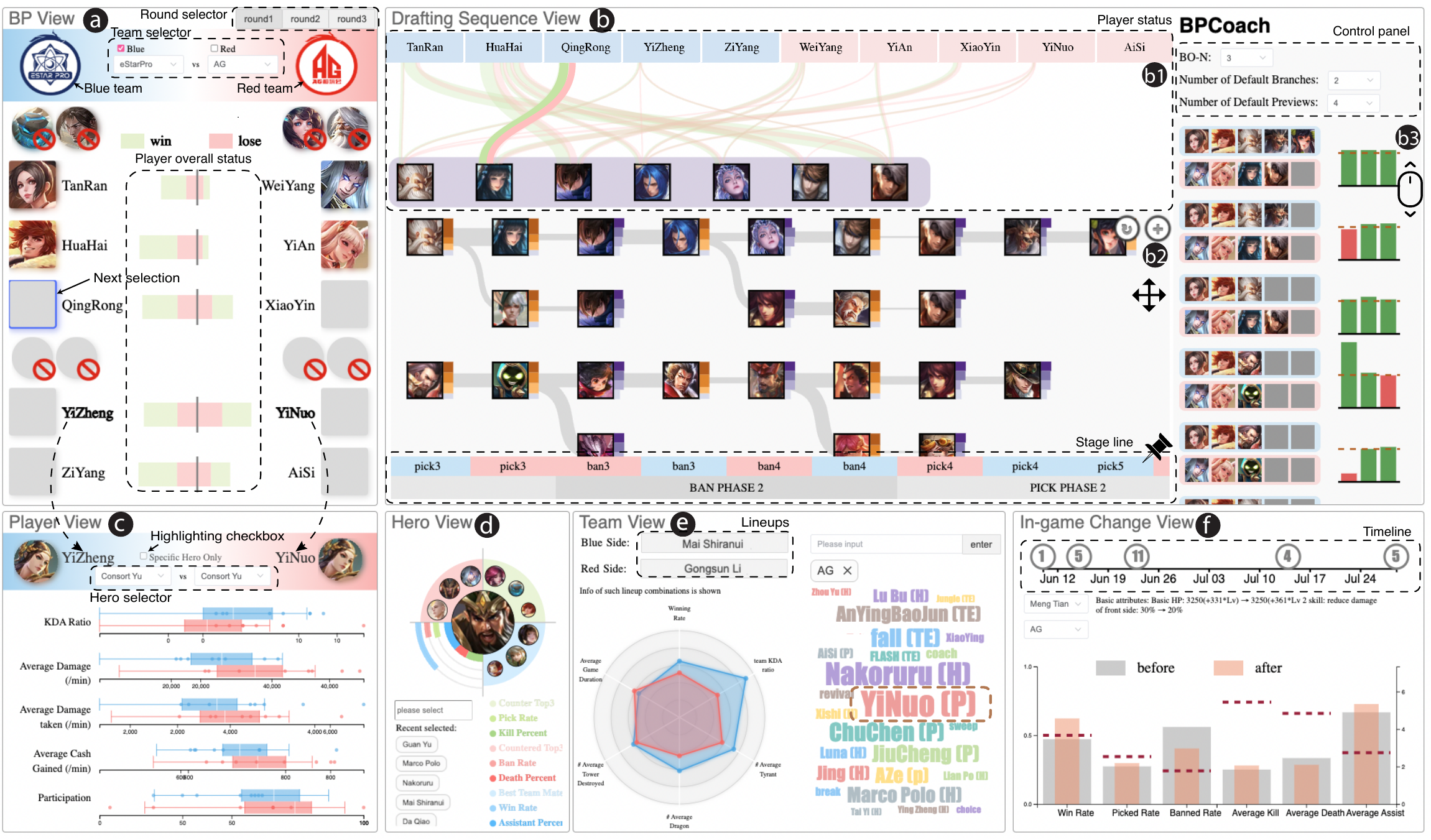}
  \caption{\textit{BPCoach} Interface. (a) \textit{BP view} supports the users to control the current hero drafting stage and gives overall information on the players. (b) \textit{Drafting sequence view} displays the drafting recommendation and prediction paths (b2), and shows the player status (b1) with the interested path. Different drafting can be compared within their scores(b3). (c) \textit{Player view} presents the player comparison between two teams with their specific hero data highlighted. (d) \textit{Hero view} shows the hero's information and relationships with others. (e) \textit{Team view} uses a radar chart to compare the overall and specific strengths of two teams and displays the keywords through a word cloud. (f) \textit{In-game change view} shows the timeline and the comparisons of hero status before and after the in-game updates.}
  \label{fig:system}
  \Description{}
\end{figure}

\begin{wrapfigure}{r}{0.5\textwidth}
  \includegraphics[width=0.5\textwidth]{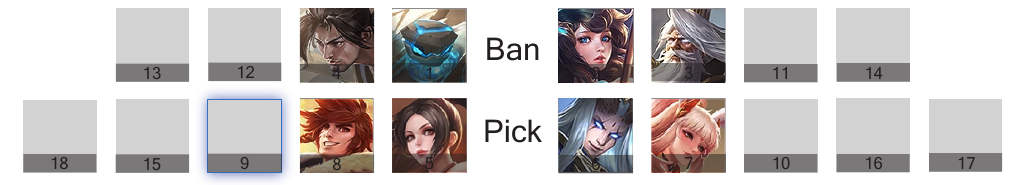}
  \caption{Design alternatives of \textit{bp view}.}
  \label{fig:notint}
  \Description{}
\end{wrapfigure}

\textit{BPCoach} enhances familiar visual metaphors to facilitate efficient comprehension of the analysis results for domain experts. We strictly adhere to the mantra of ``overview first, zoom and filter, then details-on-demand~\cite{Shneiderman2003-fn}'' to assist experts in comparing various drafting strategies and exploring the data related to hero drafting games. Based on these principles and the previously mentioned requirements, we have designed and developed the front end of \textit{BPCoach}, as shown in Fig.~\ref{fig:system}. The front end comprises five main views: the \textit{bp view} enables the user to control the current drafting process in the game and displays an overview of the player's status (\textbf{F1, F3}, \textbf{DR5}); the \textit{drafting sequence view} shows and compares possible drafting sequence paths, their alternative and customized paths, as well as player-specific conditions in the path of interest (\textbf{F1}, \textbf{DR1--3}); the \textit{player view} shows the comparisons between players from different teams in terms of overall and specific strength (\textbf{F2, F3}, \textbf{DR5}); the \textit{hero view} displays the basic information of the hero and its synergic and countering relationships with others (\textbf{F7}, \textbf{DR4}); the \textit{team view} reveals the team strengths of different drafting through a radar chart and extracts team-related keywords using a word cloud (\textbf{F4, F5}, \textbf{DR5}); the \textit{in-game change view} compares the data on the hero before and after the in-game updates (\textbf{F6}, \textbf{DR6}).

\subsubsection{BP View}
\label{sec:bp}

The \textit{bp view} (Fig.~\ref{fig:system} (a)) serves two primary functions. Firstly, it allows experts to select the blue-side and red-side teams and control the drafting progress of the match. Once the two opposing teams are determined, experts can choose the desired round and select the established heroes by clicking on the gray image, following the progress of the draft.

\begin{wrapfigure}{r}{0.5\textwidth}
  \includegraphics[width=0.5\textwidth]{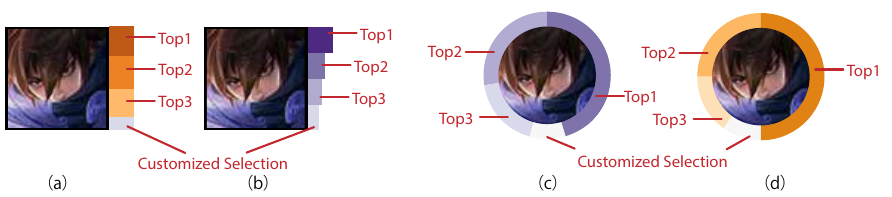}
  \caption{Sequence node design: (a) the node with a prediction next stage; (b) the node with a recommendation next stage; (c) design alternative for the node with a prediction next stage; (d) design alternative for the node with a recommendation next stage.}
  \label{fig:sequence_glyph}
  \Description{}
\end{wrapfigure}

\par The design alternative of \textit{bp view} is shown in Fig. ~\ref{fig:notint}. This design displays the heroes banned and picked by both teams horizontally on each side, and the ``ban''/``pick'' order is written on their avatars. Although this also demonstrates the ``ban''/``pick'' process, collaborating experts believe that this design is harder to adapt to. They chose the design shown in Fig.~\ref{fig:system} (a), which resembles the control panel used in tournaments. During the ``ban'' stage, a circular icon with a ``banning'' symbol is displayed. The subsequent hero selection is indicated by a blue stroke around the image. Additionally, a comparison of the overall player status from both teams is provided. In the middle of the interface, a bar chart presents information on the number of times each player has been on the field, with green and red colors representing victories and defeats, respectively. This visual representation allows experts to quickly compare the experience and overall strength of the players.

\subsubsection{Drafting Sequence View}
The \textit{drafting sequence view} consists of four components that aim to provide hero drafting recommendations and predictions and facilitate comparisons between different drafting strategies. Experts can first identify and select the paths they are interested in in the \textbf{\textit{sequence path view}} (Fig.~\ref{fig:system} (b2)). Then, they can view the corresponding player status information for the heroes on the path in the \textbf{\textit{player status view}} (Fig.~\ref{fig:system} (b1)). By clicking the interesting paths, experts can pin the paths in the \textbf{\textit{drafting comparison view}} (Fig.~\ref{fig:system} (b3)) so that drafting influence on the rest of the rounds can be obtained. Experts can trade off between recommendations, player status, and future impact.

\par In the \textbf{\textit{sequence path view.}} (Fig.~\ref{fig:system} (b2)), the sequence path is generated by utilizing the recommendation and prediction models of the back-end engine. This empowers experts to simulate potential future hero drafting scenarios. Depending on whether the upcoming stage belongs to our team or the opponent's team, the corresponding model is employed. Experts are given the ability to control the default branches and preview forthcoming steps through interaction with the control panel. This interactive feature enables experts to explore different options, make informed decisions, and strategize effectively based on the generated sequence path.

\par The generated paths are displayed as shown in Fig.~\ref{fig:system} (b2) on an infinite canvas that supports panning operations. A stage line is pinned at the bottom, indicating the stages of the drafting process. The sequence path and alternatives are encoded into a modified tree diagram. Specifically, the top path is always in a horizontal layout. The width of the paths is proportional to the score or likelihood of the child nodes. Nodes are mainly encoded as avatars of the output heroes because our experts suggest that avatars are more readable. In addition, we attach a stacked chart or bars to the right side of the avatar, as shown in Fig.~\ref{fig:sequence_glyph} (a)(b). Fig.~\ref{fig:sequence_glyph} (a) represents the node whose next stage is the prediction stage. We display the three top predicted alternative heroes with their prediction percentages stacked vertically. Similarly, Fig.~\ref{fig:sequence_glyph} (b) represents the node whose next stage is the recommendation stage. To intuitively show the recommendation scores, we use a horizontal bar chart. In both nodes, the expert can hover over the stack and the bar graph to get alternative heroes and click to expand the branch if interested. We provide experts with a customized selection to select their desired hero.

\begin{wrapfigure}{r}{0.4\textwidth}
  \includegraphics[width=0.4\textwidth]{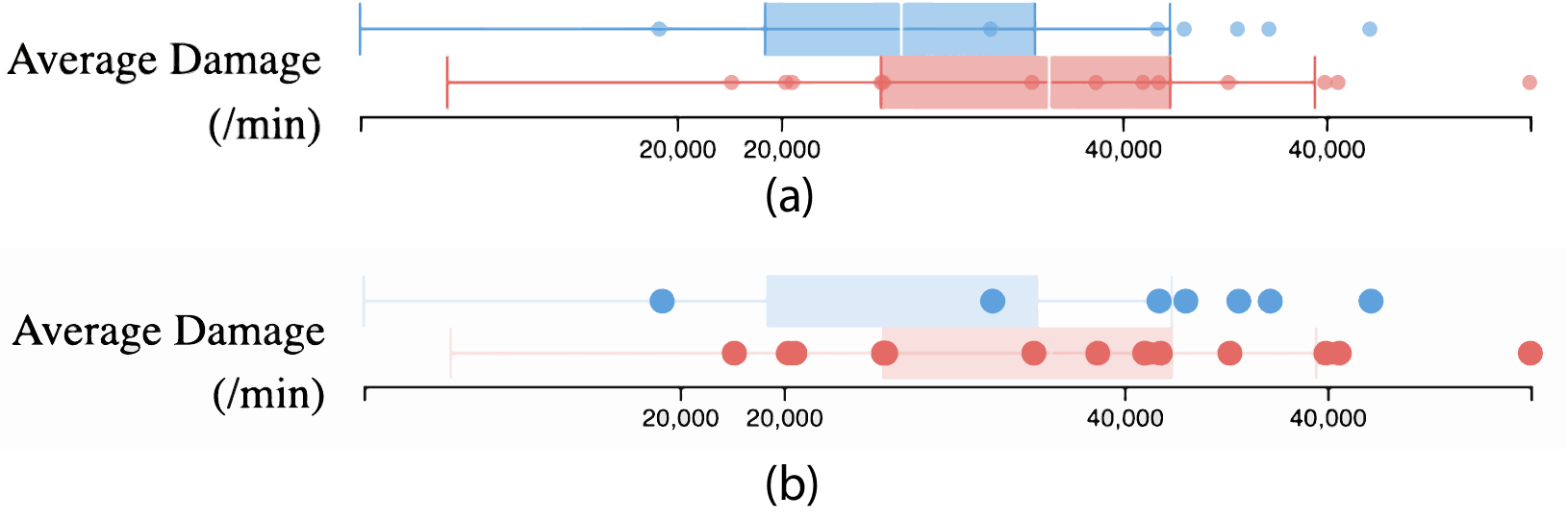}
  \caption{An example box plot in the \textit{player view}: (a) Player data on overall strength; (b) Player data with specific hero highlighted.}
  \label{fig:player}
  \Description{}
\end{wrapfigure}

\par During the design iteration, we proposed an alternative design using pie charts (Fig.~\ref{fig:sequence_glyph} (c)(d)). However, the pie chart is less intuitive in representing the status of the ``next'' stage and the colors around the avatar can be overwhelming. Therefore, we opted for the current design.

\par In the \textbf{\textit{player status view.}} (Fig.~\ref{fig:system} (b1)), after experts have selected the paths of interest, the player status view allows them to assess how the player is performing with those selected heroes in the path.

\par Players are connected to the heroes they have previously used. The width of the path indicates the number of times the player has used that hero, while the color indicates the game result associated with that hero. This allows experts to quickly determine a player's familiarity with a particular hero. While a hero may have a high drafting win rate, if the player does not perform well with that hero, the expert may consider suggesting a second choice.

\par The \textit{\textbf{drafting comparison view}} (Fig.~\ref{fig:system} (b3)) is used to compare the impact of various drafting on the rest of the rounds.

\par When the user selects an interesting path, the corresponding drafting for both sides is displayed in the comparison view as quick notes, allowing the user to easily recall their selections for that particular path. Additionally, the predicted win rate for the remaining rounds is represented as a bar chart. A dashed line is used to indicate the $50\%$ win rate threshold. Bars are colored red when the win percentage falls below $50\%$. This view provides a more intuitive comparison between two or more interesting draftings, enabling experts to make better decisions regarding which path to follow.

\subsubsection{Player View}
\label{sec:player}
Experts can select two players from different camps who have the same role in the team. Then the \textit{player view} (Fig.~\ref{fig:system} (c)) displays a more detailed view of the player's performance in the current season. 

\par Five box plots display the following player statistics: KDA ratio, average damage per minute, average damage taken per minute, average cash gained per minute, and participation rate. These box plots allow for a comprehensive comparison of the players' overall strengths in these five aspects. Additionally, experts have the option to assign a specific hero to a player and evaluate their performance on that hero. The box plot dots represent games in which the player used the assigned hero. By enabling the highlighting option at the top, as shown in Fig.~\ref{fig:player} (b), these dots can be highlighted.his feature enables experts to not only assess the overall strength of a player but also evaluate their performance on a specific hero, aiding in decisions related to whether to ``ban'' or ``pick'' that particular hero.
\subsubsection{Hero View}
The \textit{hero view} (Fig.~\ref{fig:system} (d)) shows mainly synergies and confrontations between heroes, as well as basic information about the heroes.

\par
\begin{wrapfigure}{r}{0.5\textwidth}
  \includegraphics[width=0.5\textwidth]{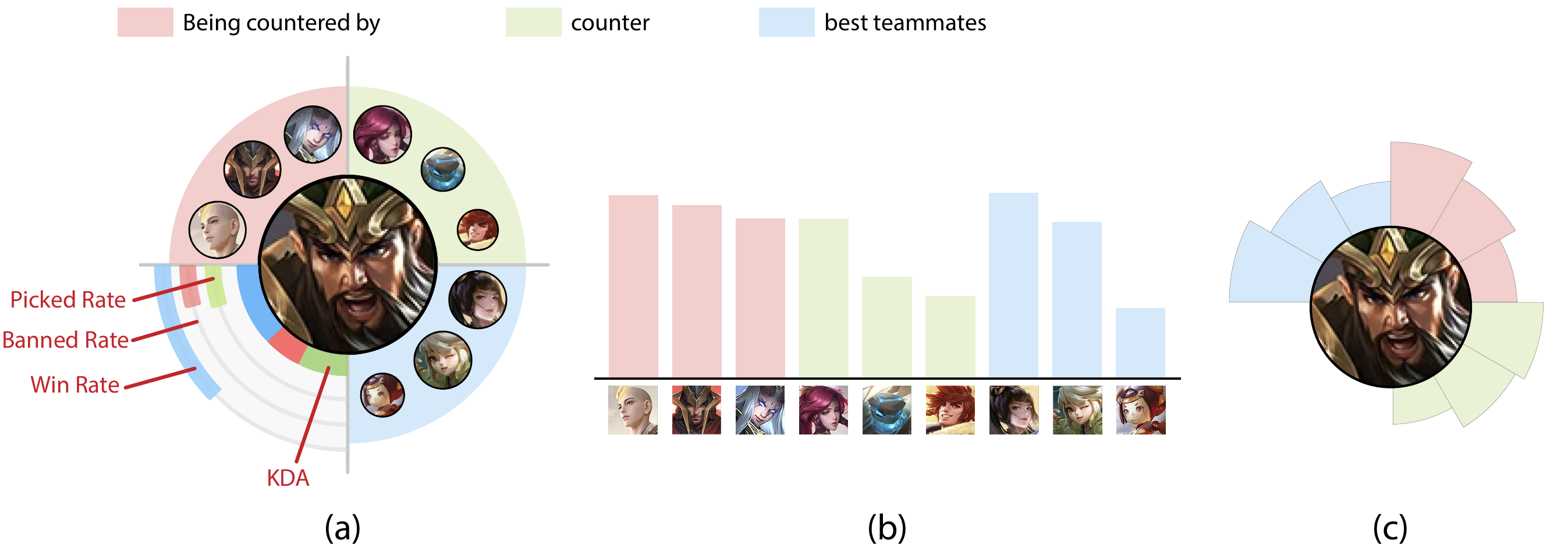}
  \caption{Hero glyph design: (a) Hero glyph showing basic hero information and synergy and antagonistic relationships between heroes; (b)(c) Design alternatives for hero glyph.}
  \label{fig:glyph}
  \Description{}
\end{wrapfigure}

 After selecting an interesting hero, the glyph of that hero is displayed and the selection is recorded in history to select it again easily. In the glyph (Fig.~\ref{fig:glyph} (a)), we divide the space around the avatar into four parts. Three of them are used to express the relationships with other heroes. The red section shows the top three heroes that the hero is countered by; the green section shows the top three heroes that the hero counters; and the blue section shows the top three heroes that are the best teammates of the hero. The radius of the avatar is determined by the influence of that hero. Meanwhile, the bottom left section contains the basic information about that hero. The innermost ring is the \textit{KDA ratio}; green indicates ``kills'', red indicates the ``deaths'', and blue indicates ``assist''. The three outer rings represent the picked rate, banned rate, and win rate, respectively. 

\par In the design iteration, we considered both bars and circular bars. As shown in Fig.~\ref{fig:glyph} (b), while the bar chart can show the influence of other heroes by the height of the bars, it lacks basic information about the hero, whereas our experts said that \textit{``basic information about the hero is very helpful to consider hero strength''}. In addition, experts pointed out the importance of direct readability of the hero's identity. Therefore, since the circular bar chart (Fig.~\ref{fig:glyph} (c)) does not directly identify the hero, it requires the user to spend more time hovering over the chart. Therefore, we finally decided to use the current design as it helps experts to absorb relationships and essential information efficiently.

\subsubsection{Team View} 
\label{sec:team}
In addition to individual performances, teamwork also plays a key role. Teams possess varying proficiencies in certain strategies that necessitate close cooperation and, as a result, pose distinct challenges to other teams.

\par The \textit{team view} (Fig.~\ref{fig:system} (e)) contains a radar chart and a word cloud that reveals the teams' strengths and play styles. The radar chart shows the team's status in six areas, including win rate, team KDA ratio, average game duration, average tyrants, average dragons, and average towers destroyed during the game. Experts can assign specific drafting to different teams with two multi-selectors. The radar chart is updated according to the selected drafting. If a team appears aggressive with a high KDA, win rate, and other indicators, the expert will consider drafting against that drafting. Word clouds are generated from one or more user-defined inputs. Experts can enter any word related to a team, player, or hero. The word cloud displays the relevant keywords around the inputs with different weights so that semantic information can be extracted. For example, when the team name \textit{``AG''} is entered, the maximum output is its player name \textit{``YiNuo''}, indicating that the player is the core of the team. Therefore, he may be a breakthrough.

\subsubsection{In-game Change View} 
\label{sec:in-game}
The \textit{in-game change view} (Fig.~\ref{fig:system} (f)) provides the changes updated by the game operator, as well as the following hero drafting changes.

\par The timeline showcases the number and date of in-game changes. By clicking on it, users can select one of the altered heroes and examine their current status. The change details are presented in a bar chart format, where encoded information includes a comparison of various metrics such as the hero's win rate, picked rate, banned rate, average kills, average deaths, and average assists before and after the update. These variations serve as indicators of the impact of in-game changes on hero drafting. Additionally, we provide specific data for a particular team, represented by a dashed line, showcasing the performance of their heroes. By analyzing these comparisons, experts can reassess the priority of certain heroes.

\subsubsection{Interactions Among the Views}
\label{sec:interaction}
\begin{wrapfigure}{r}{0.4\textwidth}
  \includegraphics[width=0.4\textwidth]{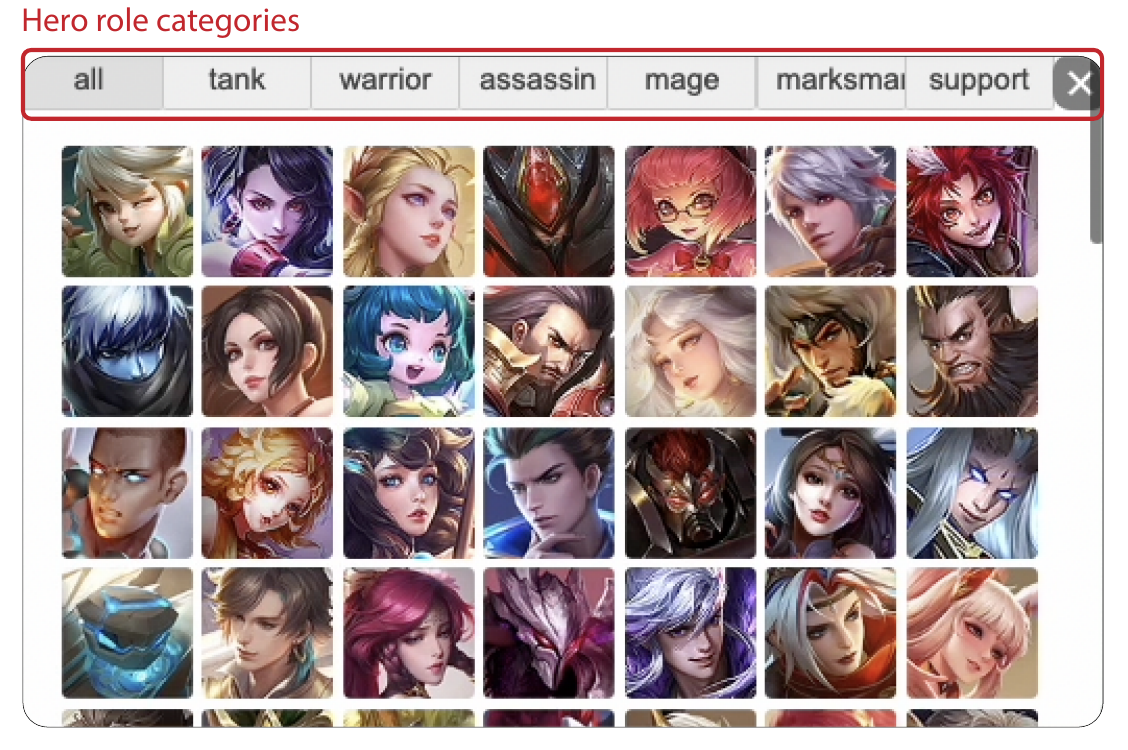}
  \caption{Hero selector in \textit{BPCoach}.}
  \label{fig:selector}
  \Description{}
\end{wrapfigure}
Rich interactions are integrated into \textit{BPCoach} to catalyze efficient in-depth analysis. (1)\textit{Panning and Scrolling:} Users can pan the canvas of sequence paths to address scalability issues when there are too many branches and steps. The drafting \textit{comparison view} can be scrolled to keep more interesting drafting. (2)\textit{Linking:} Users can click on a player name to change the player selection in the \textit{player view}, as this is a more intuitive way than presenting a selector. (3)\textit{Parameter Editing:} Users can customize their drafting path in the \textit{sequence path view}, select their drafting combinations in the team radar chart, and enter any word in the team word cloud. These interactions help with ``what-if'' analysis. (4)\textit{Hero Selector:} The hero selector in \textit{BPCoach} is built based on user habits. According to experts, it is more difficult to read literalized names than their avatars. Therefore, we encode the heroes as their avatars and sort them alphabetically as shown in Fig.~\ref{fig:selector}. Also, since there are up to $110$ heroes in the showcased game, categorizing and filtering the heroes based on their roles in the game makes the selector more familiar to the user and becomes more efficient. (5)\textit{Highlighting} In the \textit{player view}, users can check the highlighting option to highlight specific dots. Also, both the team radar chart and the \textit{player status view} can be highlighted by hovering over them. (6)\textit{Tooltips:} If we hover over each chart, the exact data including ratio, time, rate, etc. will be displayed by a tooltip, providing detailed information.

\section{Evaluation}
We evaluated the effectiveness of \textit{BPCoach} in multiple ways from the visualization community~\cite{Lai2008-br,Isenberg2013-xs}. First, we described two case studies (\textbf{RQ3, RQ5}) with our domain experts (E1-E3) who had participated in our user-centric design process. Second, we invited $12$ participants who had no exposure to our system and conducted a user study to further assess the efficacy of \textit{BPCoach} (\textbf{RQ4}). Finally, we conducted expert interviews about their user experience and feedback (\textbf{RQ4, RQ5}).

\subsection{Case I: Obtaining Hero, Player, and Team Information for Pre-match Preparation}
In the initial scenario, the experts aimed to be well-prepared for the first round of drafting. E2 stated, ``\textit{Unlike the subsequent rounds, which are fraught with variables, drafting in the first round is planned under the best-of-$N$ rule.}'' In this particular setup, the experts were positioned on the blue side, and their objective was to determine the core elements of the drafting process during the first round.

\begin{figure}[ht]
  \centering
  \includegraphics[width=\linewidth]{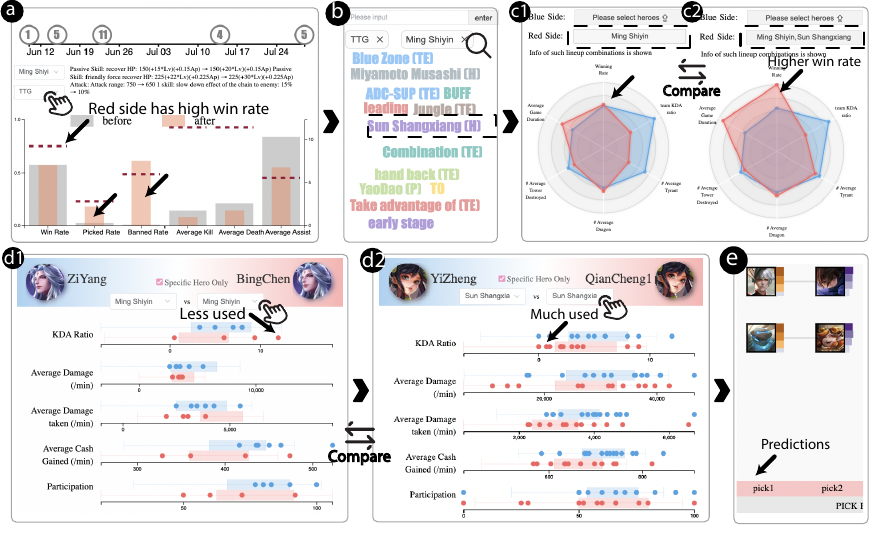}
  \caption{Case I: (a) Observing the influence of the in-game changes. (b -- c) Exploring the team strength. (d) Checking the player's strength. }
  \label{fig:case1}
  \Description{}
\end{figure}

\subsubsection{Observing the Effects of In-game Changes}
In this process, E2 first examined the impact of in-game updates in the in-game change view (Fig.~\ref{fig:case1} (a)). He found that the hero ``\textit{Ming Shiyin}'', who is a supporter, became so popular that it was rarely banned before, but now reached a banned rate of over $50\%$. In addition, the average win rate keeps over $50\%$. He was then wondering about how the red side performs after the in-game change. E2 clicked the team selector and chose to display the data of both the blue side and the red side. He admitted that both teams are good at using ``\textit{Ming Shiyin}'' and have a relatively high win rate compared with the average. At the same time, the high pick rate indicated that the opponents might have a strong tendency in picking ``\textit{Ming Shiyin}''. Therefore, the expert decided to revolve around ``\textit{Ming Shiyin}'' in the first round.

\subsubsection{Exploring the Team Strength}
To get detailed information about ``\textit{Ming Shiyin}'', E2 started to check the team strength with the team view. He searched for the red side with ``\textit{Ming Shiyin}'' in the word cloud (Fig.~\ref{fig:case1} (b)) and found a keyword called ``\textit{Sun Shangxiang}'', a marksman, indicating that the red side often chooses the combination of ``\textit{Ming Shiyin}'' and ``\textit{Sun Shangxiang}''. He then moved to the radar chart and checked the strength of the red side with only ``\textit{Ming Shiyin}'' (Fig.~\ref{fig:case1} (c1)), or the combination (Fig.~\ref{fig:case1} (c2)) of ``\textit{Ming Shiyin}'' and ``\textit{Sun Shangxiang}''. He compared the six indicators in the radar chart and found that the strength of combination is outstanding with a higher win rate. Therefore, he must act to avoid his opponent from picking both ``\textit{Ming Shiyin}'' and ``\textit{Sun Shangxiang}''.

\subsubsection{Checking the Player's Strength}
The expert proposed two methods: ``banning'', or ``picking'' one of the two heroes for his side. He then selected the corresponding players and filtered their data with ``\textit{Ming Shiyin}'' (Fig.~\ref{fig:case1} (d1)) or ``\textit{Sun Shangxiang}'' (Fig.~\ref{fig:case1} (d2)) in the player view. With the data shown, he found that the red side performance of ``\textit{Ming Shiyin}'' is not stable, while they have chosen ``\textit{Sun Shangxiang}'' much more often than ``\textit{Ming Shiyin}'' and have several outstanding performances. Therefore, E2 judged that the key to the red side lay on ``\textit{Sun Shangxiang}''. As ``\textit{Sun Shangxiang}'' of the blue side was also outstanding, E2 decided to pick ``\textit{Sun Shangxiang}'' in the first ``pick'' stage to avoid the opponent picking her.

\subsubsection{Predicting the Opponent's Selections} 
E2 generated the predictions on the following opponent's ``pick'' stages in the sequence view (Fig.~\ref{fig:case1})(e). He found that the opponent had less tendency in picking ``\textit{Ming Shiyin}'' in the next two stages. The reason might be the opponent want to leave ``\textit{Ming Shiyin}'' to the round that they could pick ``\textit{Sun Shangxiang}''. Thus, by picking ``\textit{Sun Shangxiang}'' first, E2 could still get ``\textit{Ming Shiyin}'', who is popular and strong now, in some subsequent stages.

\par With \textit{BPCoach}, experts can extract enough information in the first round to decide on the general direction. ``\textit{I can move smoothly to the view I want rather than clicking through many website pages to check what I want.}'', said E2.

\subsection{Case II: Deal with Tough Situations on the Field}
 \begin{figure}[ht]
  \centering
  \includegraphics[width=\linewidth]{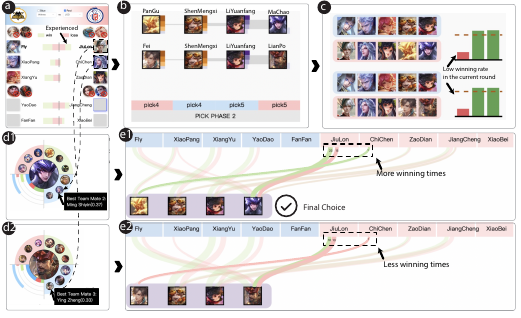}
  \caption{Case II: (a) Current drafting stage. (b - c) Obtaining recommendations from sequence path view and drafting comparison view. (d) Checking the hero information. (e) Checking the player status.}
  \label{fig:case2}
  \Description{}
\end{figure}

In the second case, E2 simulated real drafting to test if the system can help solve difficult situations. In this setting, E2 is on the red side and has to perform the last four stages, which are shown in Fig.~\ref{fig:case2} (a). The red side still needed a juggle hero and a top-lane hero.

\subsubsection{Obtaining Recommendations in Sequence View} 
E2 first generated two recommended paths in the sequence view (Fig.~\ref{fig:case2} (b)). The results suggested that regardless of choice, the opponent would like to pick ``\textit{Shen Mengxi}'' and ``\textit{Li Yuanfang}'', and E2 agreed with this prediction. He then checked the win rates for both paths and found himself in a tricky situation, as shown in Fig.~\ref{fig:case2} (c). ``\textit{Win rate is not everything, because player power cannot be ignored.}'', said E2. Player performance is crucial to the output of the game. Therefore, E2 should make relatively appropriate choices for his team so that the players can make up for the shortcomings of the drafting with their outstanding individual abilities.

\subsubsection{Checking the Hero Information}
After glancing at the player's overall status in Fig.~\ref{fig:case2} (a), E2 decided to choose a better top-lane hero for the top-lane player, \textit{JiuLon}, since he was more experienced than the juggle player, \textit{Chichen}. E2 checked the glyphs of the two heroes (Fig.~\ref{fig:case2} (d1)(d2)) of ``\textit{Ma Chao}'' and ``\textit{Lian Po}'' (both top-lane heroes), and found that both heroes have the best teammate appearing in the existing drafting and ``\textit{Ma Chao}'' has a slightly higher win rate. E2 thought both paths are relatively promising but still felt tangled.

\subsubsection{Checking the Player Status} 
E2 started checking how well the players performed with these heroes. He clicked both two paths and generated the player status views. He found that both players performed better with the heroes on the first path than on the second. Specifically, \textit{JiuLon} played ``\textit{Ma Chao}'' $38$ times and won $23$ of them, while only player ``\textit{Lian Po}'' for $23$ times and lost more than half of them. Therefore, he planned to draft along the first path and felt confident in that choice.

\par After the simulation, the experts concluded that the system is informative and can help make better, more confident choices. ``\textit{The data is reliable, so the help is meaningful.}'', said E2.

\subsection{User Study}
In professional drafting scenarios, there is no standard dataset or process for evaluating drafting recommendations. Therefore, we conducted a user study focused on evaluating the \textbf{informativeness}, \textbf{decision-making}, and \textbf{usability} of \textit{BPCoach} (\textbf{RQ5}).

\subsubsection{Participants} 
We recruited $12$ participants ($6$ female, $6$ male, $age_{mean} = 22.7$, $age_{sd} = 3.3$) via word-of-mouth. They included $3$ professional players from the collaborative team and $9$ students from a local university, all of whom had extensive \textit{HoK} experience (they had reached the top three tiers). $9$ of the $12$ had been following \textit{KPL} for over a year. We selected participants with some background knowledge because the drafting process is complex and they could provide us with more meaningful feedback. None of them had tried the proposed system before the interview. $9$ of them used the official match data website~\cite{TencentKPL} to acquire information in \textit{KPL}, while the other $3$ did not care about this information. Participants were asked to think aloud during the process. Upon completion, each participant received a $\$20$ stipend in appreciation for their contribution. 

\subsubsection{Tasks} 
In our study, each participant was given four tasks to complete. The objective of \textbf{Tasks 1-3} was to assess whether the system could provide sufficient information to facilitate professional drafting preparations. In \textbf{Task 1}, participants were asked to evaluate the performance of players throughout one season. \textbf{Task 2} required participants to assess how well a player performed on a specific hero. \textbf{Task 3} aimed to gauge participants' ability to determine the strength of a team based on their drafting choices. As for \textbf{Task 4}, the goal was to investigate whether the system could assist users in making effective drafting decisions. Participants were assigned an ongoing game scenario and were required to make their next drafting choice. Given that, not all participants were professional players, our primary focus was on examining the overall functionality and usability of the system design.

\subsubsection{Procedures} 
We compared \textit{BPCoach} with the official match data website as a baseline system where current professional teams get the information they need. The experiment consisted of four sessions. In the first session, we conducted a tutorial session for approximately $20$ minutes where we briefly introduced two systems. We then asked participants to operate these systems themselves and to ask us questions. When participants were ready, we assigned each of the four tasks described above using two separate systems. Participants were first randomly assigned to one of the systems. After completing the tasks, they were given a questionnaire containing questions on the Likert scale criteria~\cite{OBrien2018-rq}, with $1$ representing ``strongly disagree'' and $7$ representing ``strongly agree''. The questions included four items on \textit{informativeness}, two items on \textit{design convenience}, and two items on \textit{system usability}, which are shown in Table~\ref{tab:question}. Then, participants were then assigned to another system and were asked to complete the same four tasks and the questionnaires again. In addition, they were asked how they felt about the two systems.

\begin{table}[ht]
\caption{The questions in the questionnaire consisted of three parts: informativeness (Q1-Q2), decision-making (Q3-Q5), and system usability (Q6-Q8).}
\label{tab:question}
\scalebox{0.85}{\begin{tabular}{@{}cccc@{}}
\toprule
Q1 & \multirow{2}{*}{\textbf{Informativeness}}  & Accessibility                              & The information needed to plan a drafting path is easy to access. \\
Q2 &                                   & Sufficiency                                & The information is sufficient to plan the drafting.               \\\midrule
Q3 & \multirow{3}{*}{\textbf{Decision-making}}  & Comprehension                              & I can explore drafting process comprehensively.                   \\
Q4 &                                   & Assistance                                 & The system is helpful for me to plan the drafting.                \\
Q5 &                                   & Confidence                                 & I am confident that I find a suitable drafting selection.         \\\midrule
Q6 & \multirow{3}{*}{\textbf{System Usability}} & Easy to learn                              & It was easy to learn the system.                                  \\
Q7 &                                   & Understandability                          & I can understand the information provided by the system.          \\
Q8 &                                   & Intuitiveness                              & The visual design is intuitive and friendly. \\
\bottomrule
\end{tabular}}
\end{table}

\subsubsection{Hypothesis} 
We propose the following hypotheses: \textbf{\textit{H1.}} \textit{BPCoach} is superior in accessibility (\textit{H1a}) and sufficiency (\textit{H1b}) compared to the baseline system. \textbf{\textit{H2.}} \textit{BPCoach} has advantages in comprehension (\textit{H2a}), confidence (\textit{H2b}), and assistance (\textit{H2c}) in helping decision-making compared to the baseline system. \textbf{\textit{H3.}} \textit{BPCoach} is easier to learn (\textit{H3a}) and easier to understand (\textit{H3b}). \textbf{\textit{H4.}} The proposed design is more intuitive and user-friendly.

\subsubsection{Results and Analysis}
\begin{figure}[ht]
  \centering
  \includegraphics[width=0.9\linewidth]{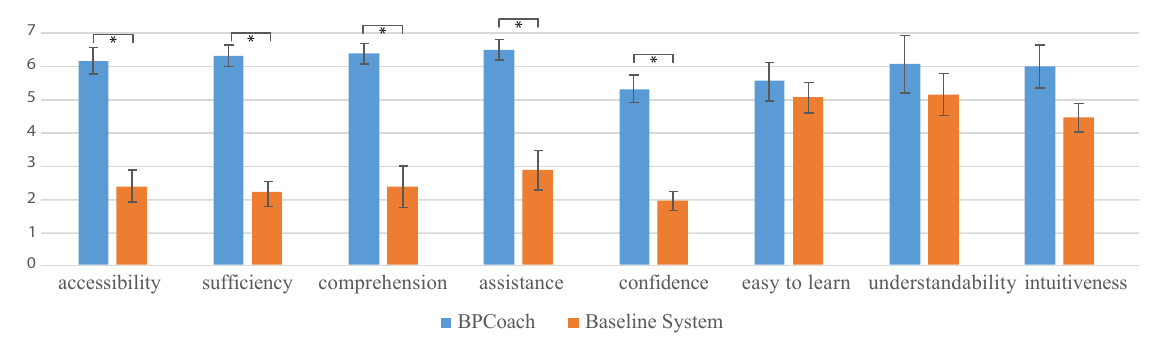}
  \caption{Means and standard errors of \textit{Baseline System} and \textit{BPCoach} in terms of \textit{informativeness}, \textit{decision-making}, and \textit{system usability} using a 7-point Likert scale. ($*: p<.05$)}
  \label{fig:result}
  \Description{}
\end{figure}

We reported quantitative ratings by the paired $t$-test. Results are shown in Fig.~\ref{fig:result} and Table~\ref{tab:result}.

\begin{table}[ht]
\caption{Paired sample t-test results.}
\label{tab:result}
\begin{tabular}{@{}cccc@{}}
\toprule
             &                   & $t$     & $Sig.$ ($2$-tailed) \\ \midrule
\multirow{2}{*}{\textbf{Informativeness}}     & Accessibility     & 8.12 & <0.001          \\
                                 & Sufficiency       & 14.20 & <0.001             \\\midrule
\multirow{3}{*}{\textbf{Decision-making}} & Comprehension     & 8.685 & <0.001          \\
                                 & Assistance        & 8.248 & <0.001           \\
                                 & Confidence        & 5.49 & <0.001               \\\midrule
\multirow{3}{*}{\textbf{System Usability}}       & Easy to learn       & 1.646 & 0.13           \\
                                 & Understandability & 0.944 & 0.36           \\
                                 & Intuitiveness     & 1.698 & 0.12           \\ \bottomrule
\end{tabular}
\end{table}

\begin{compactitem}
\item \textbf{Informativeness.} Compared to the baseline system, the proposed system scored significantly higher in both accessibility and sufficiency. In the accessibility rating, participants found it easier to access the information using the proposed \textit{BPCoach} ($Mean = 6.08$, $SD=0.90$) than the baseline system ($Mean = 2.33$, $SD=1.37$) with $p<0.05$ (\textbf{H1a supported}). In the sufficiency rating, participants found that the proposed system provided more information ($Mean = 6.25$, $SD=0.96$) than the baseline system ($Mean = 2.17$, $SD=0.84$) (\textbf{H1b supported}). One participant reported that \textit{``the distribution of players on certain heroes is meaningful and intuitive. And this kind of information is hard to get on other platforms.''} (ID:5, male, age: $18$). 

\item \textbf{Decision-making.} The proposed system scored significantly higher in all three aspects. The comprehension of the proposed system ($Mean = 6.33$, $SD=0.89$) outperformed the baseline system ($Mean = 2.33$, $SD=1.37$) within $p<0.05$ (\textbf{H2a supported}). One participant said that, \textit{``the exploration process is logical and intriguing.''} (ID:6, male, age: $28$). The proposed system ($Mean = 6.42$, $SD=0.79$) was significantly more helpful for hero drafting than the baseline system ($Mean = 2.83$, $SD=1.53$), $p<0.05$ (\textbf{H2b supported}). Participants felt more confident in their decisions made using the proposed system ($Mean = 5.25$, $SD=1.60$) than the baseline system ($Mean = 1.92$, $SD=1.16$) with $p<0.05$ (\textbf{H2c supported}). Here is representative verbal feedback, \textit{``the evaluated win rate on the rest of the rounds gives me an overview of the match. Thus, I feel more confident in picking some heroes.''} (ID:1, male, age: $24$).

\item \textbf{System usability.} Overall, these two systems received relatively high scores in three aspects. \textit{BPCoach} performs slightly better in this part. Both the proposed system ($Mean = 5.50$, $SD=1.57$) and the baseline system ($Mean = 5.00$, $SD=1.95$) were considered too easy to learn, but their differences were not significant (\textbf{H3a rejected}). The comprehensibility of both the proposed system ($Mean = 6.00$, $SD=1.54$) and baseline system ($Mean = 5.08$, $SD=1.88$) was highly rated with no significant difference (\textbf{H3b rejected}). Meanwhile, the proposed system ($Mean = 5.92$, $SD=1.73$) was rated slightly higher than the baseline system ($Mean = 4.42$, $SD=2.06$) on intuitiveness (\textbf{H4 rejected}). One of the participants said that  \textit{``the design in \textit{BPCoach} is intuitive and helps me acquire information at a glance.''} (ID:7, male, age: $18$). The reason behind this may be that both systems refer partly to the expressions that users are familiar with in the game.
\end{compactitem}

\subsection{Expert Interview}
We conducted online interviews with the aforementioned experts, E1 and E2, to collect their feedback and to observe their work process (\textbf{RQ4, RQ5}). Neither of them had tried the system prior to the interviews. First, we introduced the background of the system and conducted a 20-minute tutorial session. Then, we let them explore the system on their own and ask us questions if needed. After that, the experts were asked to perform a ``ban/pick'' simulation with the help of \textit{BPCoach}. Finally, we collected their feedback on the system's performance, visual design and interaction, generalization, and scalability.

\subsubsection{System Performance}
We asked the experts the following questions: ``\textit{Does the system meet each requirement?}'', ``\textit{Is the system sufficient to support a more effective and efficient `ban/pick' decision-making process?}'', and ``\textit{Did you encounter any potential problems in the exploration process?}'' Both experts agreed that the system was meaningful and well-organized. ``\textit{The design of the system is consistent with the logic of our exploration.}'' The recommendations provided by our system are informative and serve as a reminder for E2. E2 can iterate and optimize his drafting plan with the help of \textit{BPCoach}. In discussing the hero view, E1 said that ``\textit{new insights can be drawn from quantitatively displaying hero relationships.}'' Experts had some assumptions about some hero relationships, but they found that this is not the case.

\subsubsection{Visual Design and Interaction}
We interviewed experts in detail to get their specific comments on the visual design and interaction of each view:
\begin{compactitem}
    \item \textbf{BP view}: Both experts were satisfied with the usability of \textit{BP view} because ``\textit{its visual design aligns well with the intuition of MOBA gamers.}''
    \item \textbf{Drafting sequence view}: Experts gained some new insights into drafting selection from this perspective. ``\textit{We found some unusual patterns in the exploration. These patterns don't fit the conventional ideas based on hero strength and setting,}'' said E1. 
    \item \textbf{Player view}: A vivid box plot can show the difference between two players more visually than a coach's subjective experience and a poorly organized \textit{Excel} sheet. Both experts agreed that ``\textit{this is a good solution to our subjectivity. And box plots are actually more efficient with just a glance.}''
    \item \textbf{Hero view}: Experts appreciated the simplicity of glyphs and the wealth of information encoded. ``\textit{This helps us quickly understand the strength of the heroes in the current patch.}'' When it comes to the design, E2 stated that ``\textit{after being explained, I agree that the glyph is easy to learn. The color chosen at the background is intuitive.}''
    \item \textbf{Team view}: ``\textit{The radar plot is very useful for drafting selection, and it can provide data to support some specific tactical plans.}'' As for word clouds, ``\textit{news doesn't fully describe a team or a player, but it's a good try,}'' said E1.
    \item \textbf{In-game change view}: Experts praised this view. E2 stated that ``\textit{patch iteration has a great impact on the team's style of play. However, this dimension of data has not been previously focused on and quantified, and the bar chart is familiar.}''
\end{compactitem}

\par Both experts agreed that the designs and interactions were friendly to green hands and the system was easy to learn and use. Compared to the methods they had used before, ``\textit{we can efficiently extract some specific information instead of manually check the battles one by one,}'' said E1. ``\textit{The visualizations also help me get and understand the data faster.}'', E2 added. With \textit{BPCoach}, the experts' exploration process becomes simple and efficient.

\section{Discussion}
\subsection{Findings}
The findings are embodied in how the five research questions are answered and what can we learn from them.

\begin{compactitem}
\item \textbf{RQ1: What is the information needed for a coach's decision on hero drafting?} By performing expert interviews, we successfully extracted seven factors (F1 -- F7)~\cite{Summerville2016-ds,Summerville2016-gm,Yu2019-jk}, i.e., comprehensive coverage of factors affecting hero drafting. Since previous works on win rate prediction only considered F1 and F7, the accuracy of around $60\%$ has been unbreakable for a long time, most likely related to the absence of other factors. Therefore, future work can be done from this perspective.

\item \textbf{RQ2: What should be an acceptable approach for information processing?} Handling complex but small amounts of data is always a challenge for machine learning based methods, which leads to ignoring unavailable features or simulating an ideal scenario~\cite{Hanke2017-xi,Porokhnenko2019-bx,Chen2018-sl,Chen2021-ls}. To find out a practical approach for real application, we built a visualization system. The interactive user interface provides the opportunity for the model and humans to iterate together for optimization. Therefore, future work could integrate this ``human-in-the-loop'' approach into the proposed approach. 

\item \textbf{RQ3: What scenarios and when will this system be used?} In the study, we presented two use scenarios to verify that the system can be used before and during a match. However, currently, it is not permissible to utilize an AI system like \textit{BPCoach} in actual matches due to rule restrictions. Experts have proposed that the system could be employed during scrims (training matches) to gain fresh insights through collaboration with AI. Specifically, experts believe that in cases involving new team formations or personnel rotations{~\cite{Freeman2019-cs}}, \textit{BPCoach} can offer a rapid and meaningful solution for training and team cohesion. Therefore, for the time being, our focus is on providing a strategy for obtaining and interacting with relevant data rather than circumventing the game's rules.

\item \textbf{RQ4: What is the usability and effectiveness of the adaptive system in hero drafting?} We conducted a user study to test the usability and effectiveness of \textit{BPCoach}. The results validated the merits of the proposed system. Therefore, we can infer that this workflow can lead to an effective tool for complex decision-making problems.

\item \textbf{RQ5: How will coaches interact and collaborate with such an AI-enabled system?} Expert interviews and two cases provide us with an opportunity to see the process of iterative decision-making using the system. Primary decisions are provided by the model, and experts then start to compare the decisions and utilize other information to optimize the decision. While we cannot guarantee that every decision is the best one, we can see that the ``human-in-the-loop'' process can give experts more confidence in their decisions than a black box model. How to open the black box of a model is a question worth thinking about.
\end{compactitem}

\subsection{Contribution to Pre-game Platforms}
\label{dis:realize}
\textit{BPCoach} effectively incorporates and enhances the interactions of seven factors outlined in Table{~\ref{tab:comp}}. The predicted and recommended drafting sequences (\textbf{F1}) are based on the global ``ban''/``pick''  rule, and \textit{BPCoach} explicitly displays these potential sequences as paths in the \textit{drafting sequence view}. The player's performance on a specific hero (\textbf{F2}) and their overall status (\textbf{F3}) are presented in the \textit{player view} and \textit{bp view}, respectively. These views establish links between players to ensure coherence and consistency. Moreover, the proficiency of each player with specific heroes (\textbf{F2}) is displayed at the top of the \textit{drafting sequence view}, enabling a quick and intuitive visualization. By simply clicking on a path, users can switch between heroes in different sequences. \textit{BPCoach} summarizes team performance (\textbf{F4}) in the \textit{team view} based on various heroes and their combinations. The player style (\textbf{F5}) is also depicted in the \textit{team view}. To achieve a more personalized play style assessment, we employ a method that extracts keywords from news articles. This approach allows us to go beyond fixed labels and identify not only a player's style but also the style of a team or a specific hero, among other possibilities. The \textit{in-game change view} employs a bar chart to facilitate the easy perception of trends following game updates. The relationship among heroes is presented in the \textit{hero view}, which is linked with the \textit{drafting sequence view} to alleviate the need for repetitive clicks and streamline the user experience.

\par In summary, \textit{BPCoach} makes significant contributions in four key areas. First, we have identified and incorporated seven crucial factors into a comprehensive system, ensuring that all relevant aspects are considered. Second, we have taken a global perspective when addressing the challenge of automated prediction and recommendation. Third, we have leveraged visualization techniques to enhance the user experience and enable quick comprehension of complex data. This visual approach enables users to grasp important information at a glance. Fourth, we have implemented an interactive interface that minimizes unnecessary interactions and streamlines the user's workflow. This reduces the burden on users and ensures a smoother and more efficient experience when interacting with the system.

\subsection{Generalizability}
Our workflow and system can be applied to drafting in other games with minimal adjustments to the algorithms and interface, due to the similarity of hero drafting mechanics.

\subsubsection{Recommendation and Prediction Algorithms}
\begin{table}[ht]
\caption{Comparison of hero drafting processes in professional tournaments of different MOBA games.}
\label{tab:othermoba}
\scalebox{0.85}{\begin{tabular}{@{}cccc@{}}
\toprule
                      & ``B/P'' sequence                                                        & No. of heroes & Global ``b/p'' \\ \midrule
\textit{LoL}          & b1-b2-b1-b2-b1-b2-p1-p2-p2-p1-p1-p2-b2-b1-b2-b1-p2-p1-p1-p2             & 163              & ·              \\
\textit{DotA2}         & b1-b2-b1-b2-p1-p2-p2-p1-b1-b2-b1-b2-b1-b2-p2-p1-p1-p2-b1-b2-b1-b2-p1-p2 & 117              & ·              \\
\textit{\textbf{HoK}} & b1-b2-b1-b2-p1-p2-p2-p1-p1-p2-b2-b1-b2-b1-p2-p1-p1-p2                   & 113              & \Checkmark     \\ \bottomrule
\end{tabular}}
\end{table}

We compared the two other most famous MOBA tournaments - \textit{LoL} and \textit{DotA2}, with \textit{KPL}. As shown in Table{~\ref{tab:othermoba}}, there are slight variations in the hero sequences and numbers. These processes generally follow a specific order, consisting of ``ban'' and ``pick'' options exclusively. The number of heroes in all three games is also similar, with approximately $100$ heroes in each. Consequently, our MCTS-based recommendation model and Markov-based prediction model can be applied to all three games. Additionally, in \textit{LoL} and \textit{DotA2}, where the global ``ban''/``pick'' rule is not enforced, the computational complexity can be reduced. The reward rule in MCTS is modified to consider only the win rate of the current round, without taking into account subsequent rounds{~\cite{Chen2018-sl}}. Thus, the recommendation and prediction algorithms can generally be adapted to other MOBA games by configuring the specific heroes and the drafting order.

\subsubsection{Interactive Visual Analytics} 
The user study included three representative evaluations to verify the effectiveness of \textit{BPCoach} in drafting plans. Therefore, as long as the relevant data contains sufficient dimensions such as player and hero indicators, the system can be applied to the game. However, due to the inclusion of more complex gameplay in \textit{LoL} and \textit{DotA2}, it is highly likely that there are additional quantitative indicators to assess the abilities of players or heroes. To address this issue, we suggest incorporating more data on the charts or including additional charts to enhance the system's capabilities.

\subsubsection{Generalization to Non-professional Players}
We also took into consideration how to generalize the system for non-professional players. To cater to these players, we simplified the algorithm to focus on a one-round problem. In the visualization interface, most views can be utilized, with the exception of the \textit{team view}, as non-professional players typically do not have a fixed five-player team. However, many of them have a consistent group of teammates, forming a sub-team of duo queues or trios. In such cases, the five-player team in the \textit{team view} can be transformed into a sub-team to address this issue. Furthermore, while it may not be possible to assess the compatibility of randomly matched players, it is feasible to showcase their performance when encountering the hero that one plays. This can aid in understanding their compatibility at least with specific heroes. Another challenge in the \textit{team view} is the lack of news data for non-professional players. In this situation, label-based methods employed by commercial platforms can be valuable. However, we can go beyond providing labels solely for individual players. Similar algorithms can be used to assign labels to teams, players using specific heroes, and more. This approach allows us to refine the initially general labels across different dimensions, providing more nuanced insights. Lastly, we recognize the success of many commercial platforms, such as \textit{Senpai}, in creating in-game apps. These apps enable users to seamlessly access the platform without the need for frequent switching between the game and the platform itself. Therefore, future work can involve the development of in-game support to provide a more integrated and convenient experience for users.

\subsection{Lessons Learned}

\subsubsection{Leave Human Factors to Humans}
In \textit{BPCoach}, we only used three factors (\textbf{F1}, \textbf{F6}, \textbf{F7}) related to heroes to train the model.
We did not input all the identified factors because the remaining involves human factors related to players and teams, which are less objective and stable than hero data.
For example, a player adept at a specific hero may underperform when using this hero if they are not in top form.
However, a model built with historical player data may wrongly predict a high win rate on this player-hero pair.
For more accurate decision-making, it  requires on-the-spot observation and human intervention.
Our collaborating experts insisted that when humans are involved in these factors, the responsibility for making trade-offs should remain with humans rather than being delegated to models.

\subsubsection{Tap into Expertise}
During the iterative design process, we have gained a deep appreciation for the unique qualities that MOBA environments possess. We recognize that these qualities require specialized knowledge and expertise to design effectively. As a result, we have collaborated closely with domain experts to ensure that our design incorporates their insights and experience. Specifically, we designed the interface resembling the \textit{HoK} style, such as the interactions of the hero selector (Section{~\ref{sec:interaction}}) and the visual appearance \textit{bp view} (Section{~\ref{sec:bp}}). Additionally, experts also suggested using hero avatars instead of showing hero names solely, harnessing the power of visual cues for quicker character recognition and selection. This collaborative approach has been instrumental in developing an interface that not only adheres to MOBA-specific affordances but also enhances overall usability and user satisfaction.

\subsubsection{Considerations of AI-enabled Systems in Tournaments}
The trend of using AI in tournaments is becoming increasingly popular{~\cite{Wu2023-ym,Wang2022-ov}}, but the use of AI systems is restricted in formal matches{~\cite{Ma2022-ph}} due to fairness issues and the impact on the audience experience{~\cite{Wallace2012-bs}}. We believe that a unified platform could help ensure fairness. For example, in some recreational Go matches, both sides use a unified AI system as a teammate. Thus, if \textit{BPCoach} is used as a unified platform by both sides in MOBA games, it will not affect the fairness of the game. Furthermore, to ensure the entertainment value of such systems, it is crucial for AI to be used as an assisting tool{~\cite{Hern2017-ur}} and not as a replacement for human decision-making{~\cite{Weisz2023-qi}}. Previous work has also shown that the interaction with AI and the differences in understanding among players can add to the excitement of the game{~\cite{Munyaka2023-fy,Merritt2011-uz}}. As \textit{BPCoach} requires humans to make decisions, the entertainment value of the game will not be affected. In conclusion, the use of AI in tournaments has the potential to enhance the overall experience for players and audiences alike, but it is important to consider the impact on fairness and entertainment value. We should propose unified platforms for formal matches and use AI as an assisting tool rather than a decision-maker.

\subsection{Limitations}
There are several limitations to this work. First, we only have access to professional battle data, which is insufficient. As experts have suggested, professional data is different from ordinary data. Therefore, models with only limited professional data do not perform as well as models with millions of data. Second, since these battles are from official public matches, experts believed this leads to some bias because teams usually try more offensive drafting in scrims, but are conservative in official games. Third, running the MCTS algorithms repeatedly is time-consuming. If we limit the time, the algorithm may not have converged to a good enough answer. Therefore, the current output still has a certain randomness.

\section{Conclusion and Future Work}
In this study, we propose a visual analytics approach to exploring the hero drafting process in professional tournaments with the global ``ban''/``pick'' rule. We first extract the requirements through several discussions based on experts' opinions and finally specify the design requirements. The proposed \textit{BPCoach} helps to get information quickly, prepare for the tournament, and plan the drafting on the field. Two case studies, a user study, and expert interviews validate the efficacy of our approach. Currently, \textit{BPCoach} is only applicable to professional tournaments. We expect to extend the system to fit the non-professional players, using a more suitable model and more data, and extracting more information from the non-professional players.

\begin{acks}
We would like to express our gratitude to our domain experts and the anonymous reviewers for their insightful comments. This work is funded by grants from the National Natural Science Foundation of China (No. 62372298), the Shanghai Frontiers Science Center of Human-centered Artificial Intelligence (ShangHAI), and the Key Laboratory of Intelligent Perception and Human-Machine Collaboration (ShanghaiTech University), Ministry of Education.

\end{acks}


\bibliographystyle{ACM-Reference-Format}
\bibliography{paperpile}
    
\end{document}